\documentclass{aa}  
\usepackage{graphicx}
\usepackage{txfonts}
\usepackage{rotating}
\usepackage{amssymb}

\newcommand{\xmm}{{\it XMM-Newton}}
\newcommand{\epic}{{\it EPIC}}

\newcommand{\pn}{{\it pn}}
\newcommand{\mos}{{\it MOS}}
\newcommand{\chandra}{{\it Chandra}}

\newcommand{\chired}{$\chi^2_\nu$}

\begin{document}
   \title{XMM-Newton view of galaxy pairs: activation of quiescent black holes? \thanks{Based on observations obtained with XMM-Newton, an ESA science mission with instruments and contributions directly funded by ESA Member States and NASA.}}

\authorrunning{E. Jim\'enez-Bail\'on et al.}
\titlerunning{XMM-Newton view of galaxy pairs}
   \subtitle{}

   \author{E.    Jim\'enez-Bail\'on\inst{1}, N.  Loiseau   \inst{2},  M.   Guainazzi\inst{2},
G.  Matt\inst{1},
D. Rosa-Gonz\'alez\inst{3},  E.  Piconcelli\inst{4}, \and M. Santos-Lle\'o\inst{2}}

 \offprints{ejimenez@fis.uniroma3.it}

 \institute{Dipartimento di Fisica di l'Universita Roma 3, Via della Vasca Navale 84, I-00146 Roma, Italy
\and
XMM-Newton Science Operations Center, ESAC, ESA, Apartado 50727,
  E-28080 Madrid, Spain
\and
Instituto Nacional de Astrof\'{\i}sica Optica y Electr\'onica, Luis Enrique Erro No. 1. Tonantzintla, Puebla, C.P. 72840, M\'exico
\and
Osservatorio Astronomico di Roma (INAF), Via Frascati 33, I-00040 Monteporzio Catone, Italy
}

\date{Received September 15, 1996; accepted March 16, 1997}

 \abstract{We report on \xmm\ observations  of three nearby galaxy pairs,
AM0707-273,  AM1211-465,   and  AM2040-674.   All   six  galaxies  were
previously  classified  as  HII  galaxies  based  on  optical  and  IR
spectroscopic analysis. All galaxies  were detected with \xmm\ and each
member  was isolated  and  analyzed independently.   The X-ray  spectra
reveal strong evidence of AGN  activity in the NE member of AM1211-465
pair.       We           measured      a      luminosity      of
1.94$^{+0.11}_{-0.15}\times10^{42}$ erg/s in the 2-10 keV band and the
presence  of a  neutral FeK$\alpha$  line with  a confidence  level of
98.8\%. The high n$_{\rm{H}}$ value, $2.2\pm0.2\times10^{22}$
cm$^{-2}$, would explain the misclassification of the source. Marginal
evidence  of  AGN  nature was  found  in  the X-ray  spectra  of
AM1211-465SW  and  AM0707-273E.   The  X-ray  emission  of  the  three
remaining  galaxies  can be  explained  by  starburst activity.  

}

\maketitle

%

\section{Introduction}

Based on  optical surveys,  only about 0.1\%  of all  quasars observed
have one, or more, nearby quasar companion at the same redshift (e.g.,
Kochanek  1995; Hewett  et al.   1998).  However,  there is  a growing
number of  identifications of  true pairs of  quasars or  as excellent
candidates (see Hennawi et al.  2006 on SDSS results). Although highly
redshifted pairs tend to be interpreted as the result of gravitational
lensing by dense objects in the  line of sight, many of them cannot be
explained  in this  way,  and  even the  spectral  features that  were
supposed to indicate that they were  one and the same object have been
interpreted as a common feature of all quasars (Mortlock et al. 1999).
For  these cases  the  most reasonable  interpretation  is that  their
activity  is  the  result  of  their  interaction.   Even  though  the
interaction  of two  large galaxies  is  not always  strong enough  to
create new supermassive black holes in the cores, it is plausible that
the interaction  can refuel  the quiescent black  holes hosted  by the
majority of known galaxies.

Assuming  that most  of the  galaxies host  a supermassive  black hole
(Kormendy et al.  1997), the  key question in our understanding of AGN
formation  and evolution  would  be the  mechanisms  that trigger  this
nuclear  activity.    Galaxies  interactions are  thought  to  be
effective in driving  the gas from the outer parts  of the galaxy into
the circumnuclear  region through loss of the angular  momentum induced by
tidal  forces  (Jogee  et  al.  2005).  The inflow  of  gas  could  be
intercepted by  the starburst activity  in the circumnuclear  region or
could continue falling into the nucleus feeding the back hole (Shlosman
et al. 2005).  However, no direct causal links  between the degree of
the interaction  and the AGN activity  have been found up  to now.

Recent  X-ray   detections  show  that   many  interacting  galaxies
classified as non-AGN are in  fact active (e.g. Maiolino et al. 2003).
These   findings  prove   that  optical   spectroscopy   is  sometimes
inefficient in revealing the presence of an AGN. The X-ray study of a
small sample of  binary quasars has revealed that all  of them suffer some
degree  of  absorption  and at  least  half  of  them resulted  to  be
Compton-thick (Komossa et al. 2003;  Ballo et al.  2004; Guainazzi et
al. 2005). These findings show  that a significant number of AGN pairs
might be missed by optical  surveys due to heavy obscuration affecting
one (or both) of the members.

In this  paper we have  studied the X-ray  emission of three  pairs of
nearby interacting galaxies observed with \xmm: AM0707-273, AM1211-465,
and  AM2040-674.   In the  following  section  we  summarize the  data
analysis of  the observations.    All the studied galaxies  can be
classified as having some kind  of activity in the central regions. In
Sect.~3,  we discuss  the origin  of this  nuclear emission  as either
starburst or AGN).   The main conclusions of this  work are presented
in Sect. 4.

\section{Observations and data reduction of the \xmm\ data}

The   three  pairs  of  galaxies   (AM0707-273,  AM1211-465,  and
AM2040-674) were selected from the sample of Sekiguchi \& Wolstencroft
(1992) of interacting doubles of  comparably sized galaxies, i.e. with
relative sizes of  members to be within a factor  of two. The galaxies
also  present  a disturbed  appearance  in  optical  and/or  IR  images
proving the   interaction.   The  pairs   present  enough  projected
separation  distance  to  be  spatially  resolved by  \xmm.   All  six
galaxies in the sample are classified as HII galaxies based on optical
and IR data (Sekiguchi \& Wolstencroft 1992).


The details  of the  \xmm\ (Jansen et  al.  2001) observations  of the
pairs  of galaxies are  summarized in  Table~\ref{table:xmm_obs}.  The
raw  data  were processed  using  the  standard  {\it Science  Analysis
Systems,  SAS}, v.7.0.0.   (Gabriel et  al.  2004).  The  most updated
calibration files available  in July 2006 were used  for the reduction
of  the data.  We removed time  intervals corresponding  to high
background   using    the   method   described    in   Piconcelli   et
al. (2004).  According to the  {\it epaplot SAS} task,  no significant
pile up was detected in none of the studied sources.

\begin{table*}[htb]
\caption{General properties and details of the observations of the galaxy pairs}\label{table:xmm_obs}
\begin{tabular}{lllllll}
\hline
{\bf Target} & {\bf Obs. ID} & {\bf Obs. date} & {\bf Obs. Mode} & {\bf Filter} & {\bf \pn\ Exp.} & {\bf Net pn CR} \\
 & & & (1) & (2) & (3) & (4) \\
& & & & & ks & count/s  \\ \hline\hline
AM 0707-273 & 0300930201 &  2005-09-27 & FF & M &  \\
$\,\,${\it AM 0707-273E} &  &  &  &  &  11.33 & $(2.26\pm0.16)\times10^{-2}$ \\
$\,\,${\it AM 0707-273W} &  &  &  &  &  11.33 & $(2.99\pm0.18)\times10^{-2}$ \\ \hline
AM 1211-465 & 0300930101 &  2005-09-28 & FF & T & \\
$\,\,${\it AM 1211-465NE} &  &  &  &  &  12.88 & $(2.55\pm0.05)\times10^{-2}$ \\
$\,\,${\it AM 1211-465SW} &  &  &  &  &  5.48 & $(2.8\pm0.3)\times10^{-2}$ \\\hline
AM 2040-674 & 0300930301 &  2005-09-24 & FF & T & \\ 
$\,\,${\it AM 2040-674N} &  &  &  &  &   9.83 & $(7.1\pm1.0)\times10^{-3}$ \\
$\,\,${\it AM 2040-674S} &  &  &  &  &  12.22 & $(1.17\pm0.11)\times10^{-2}$ \\\hline
\end{tabular}\\
{  \scriptsize Notes:  (1) {\bf  Observations Modes:}  FF:  Full Frame
Mode; (2) {\bf Observations Filters:}  T: Thin, M: Medium; (3) Useful
\pn\ exposure; \\
(4) Net \pn\ count rates in the 0.2-10~keV band.}
\end{table*}

\subsection{Imaging analysis} \label{sec:imaging_analysis}

\begin{figure}

\includegraphics*[width=44mm,angle=0]{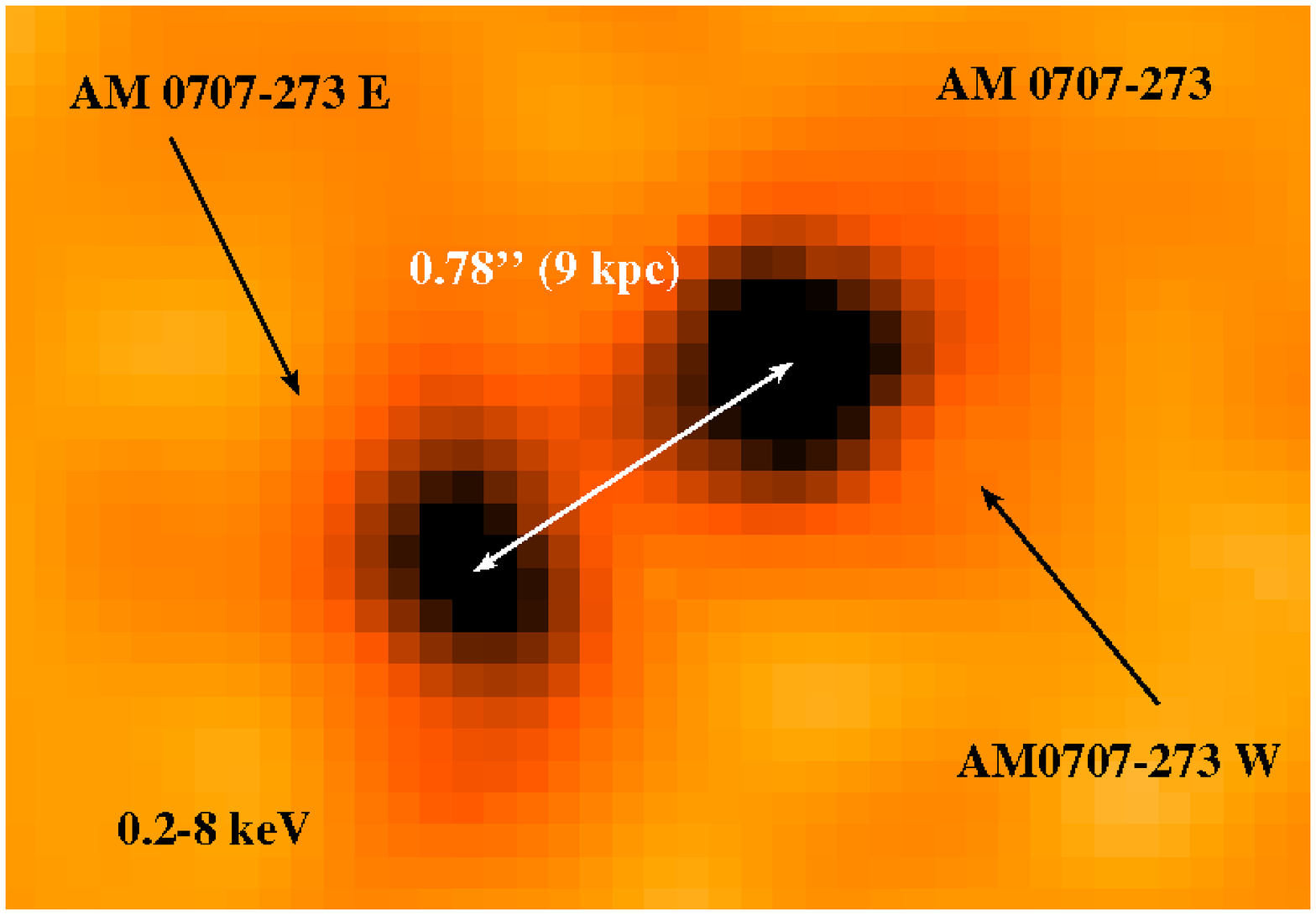}\includegraphics*[width=44mm,angle=0]{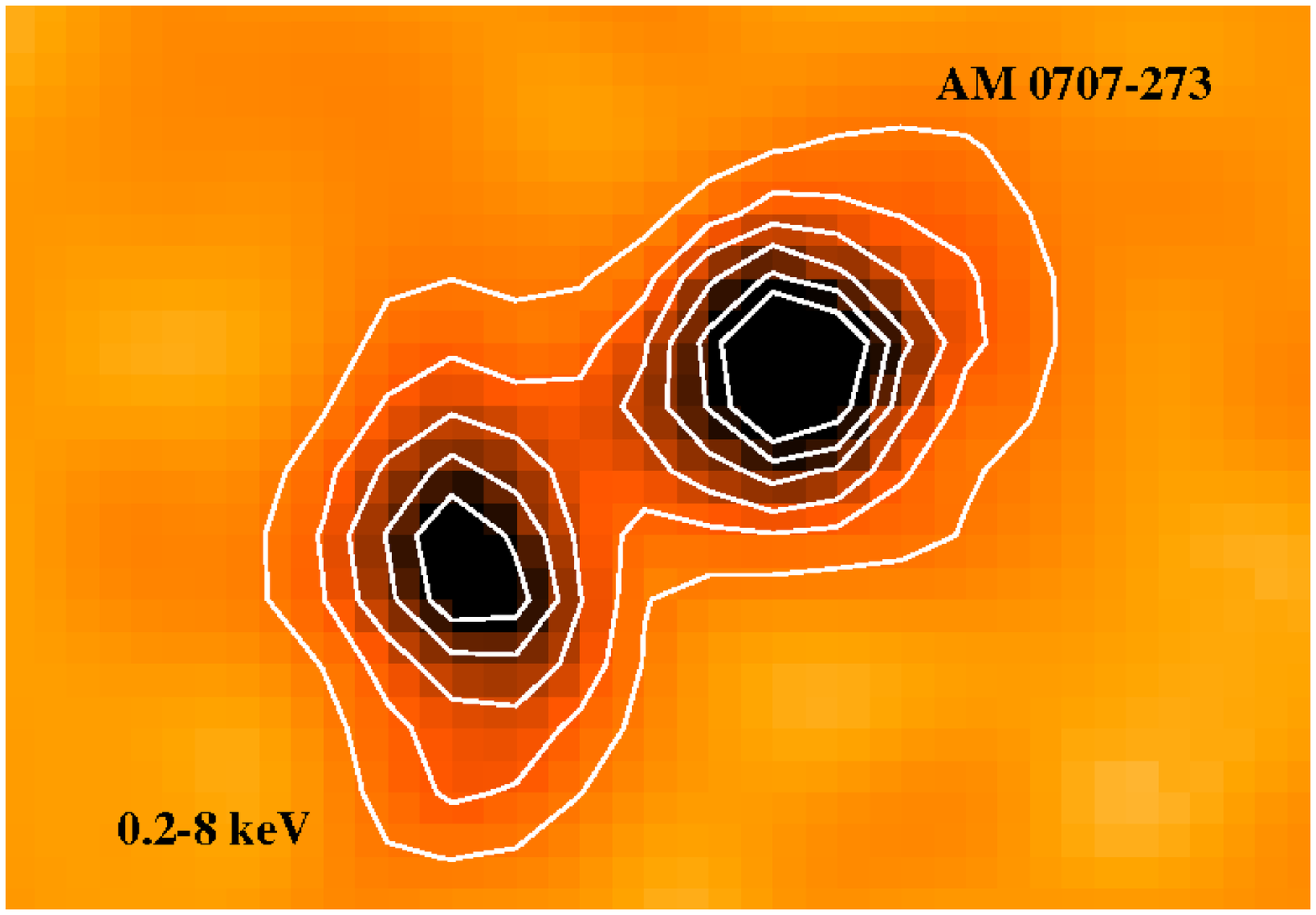}
\includegraphics*[width=44mm,angle=0]{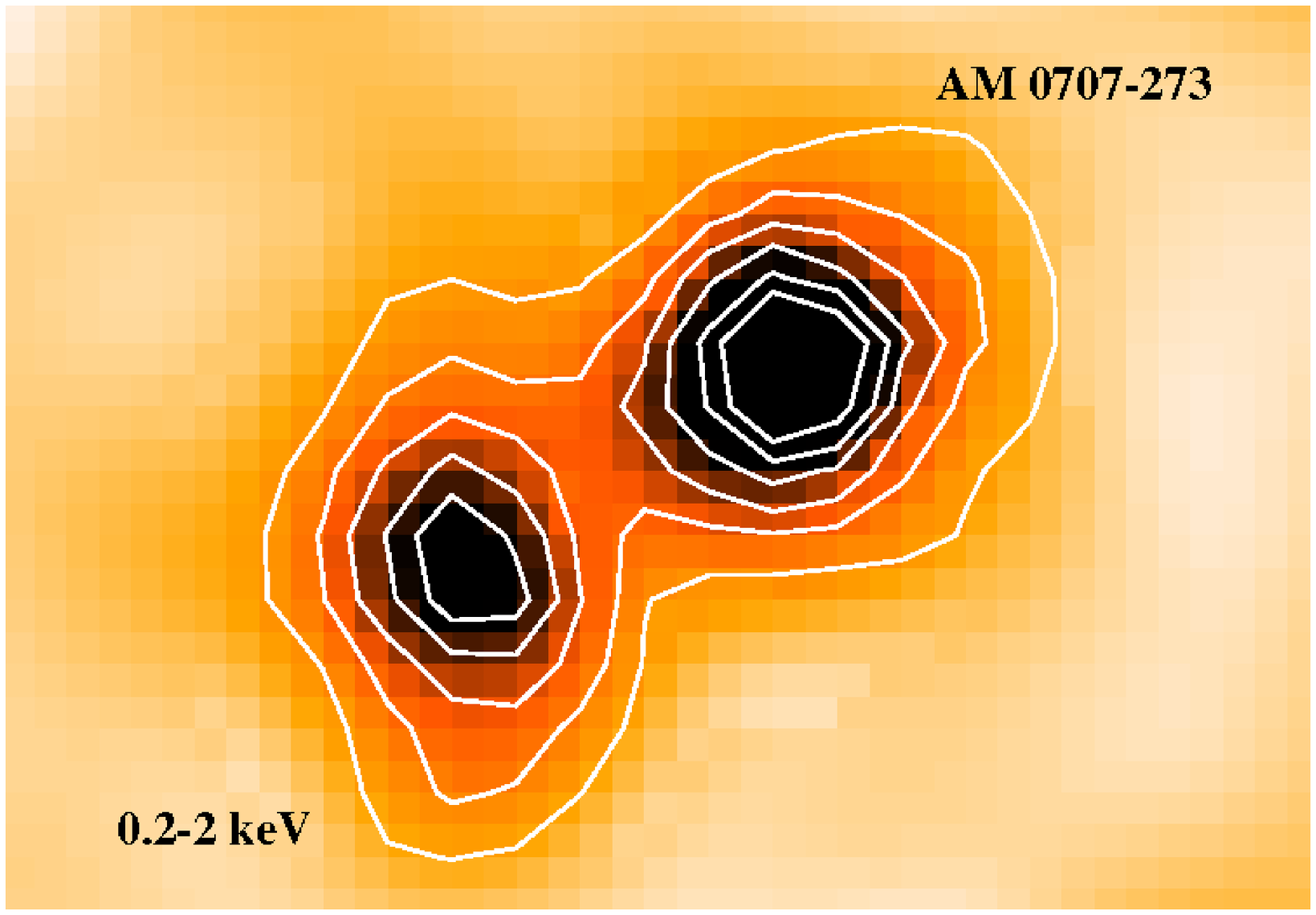}\includegraphics*[width=44mm,angle=0]{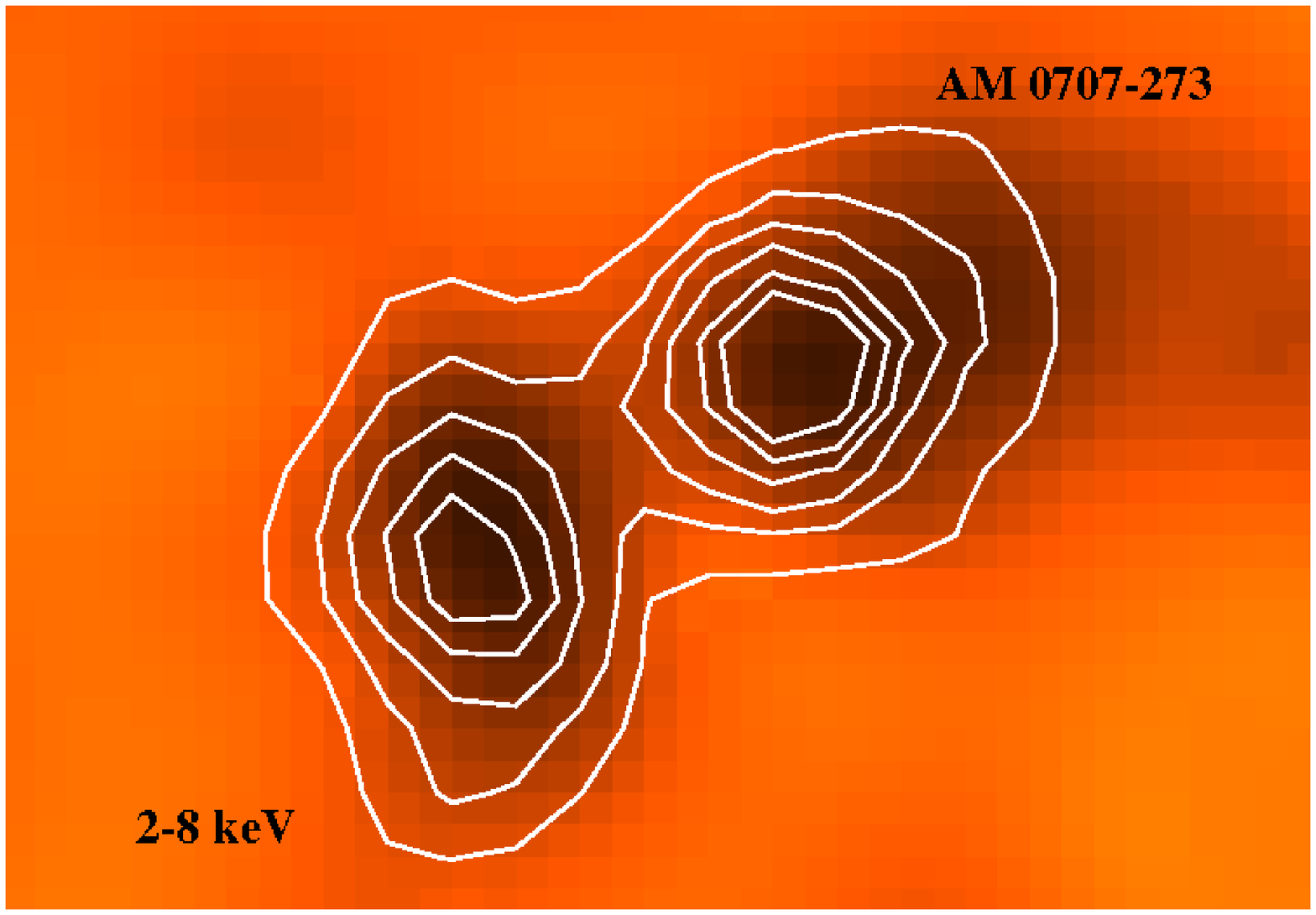}
\caption{Smoothed \pn\  images of AM~0707-273   on a logarithmic scale from 0 to 150 counts . From top  to bottom and
left  to right:  (a) Location  and separation  distance of  the sources
superimposed  on  the smoothed 0.2-8~keV  image.  (b) 0.2-8~keV  surface
brightness contours superimposed  on the
smoothed  0.2-2~keV  image. Contour levels are at 0.03, 0.2, 1,7, 45, and 285 counts. (c) 0.2-2~keV  surface
brightness contours superimposed  on the
smoothed  2-8~keV  image.   (d) 2-8~keV  surface
brightness contours superimposed  on the
smoothed  0.2-8~keV  image. }\label{fig:am0707}
\end{figure}

\begin{figure}
\includegraphics*[width=44mm,angle=0]{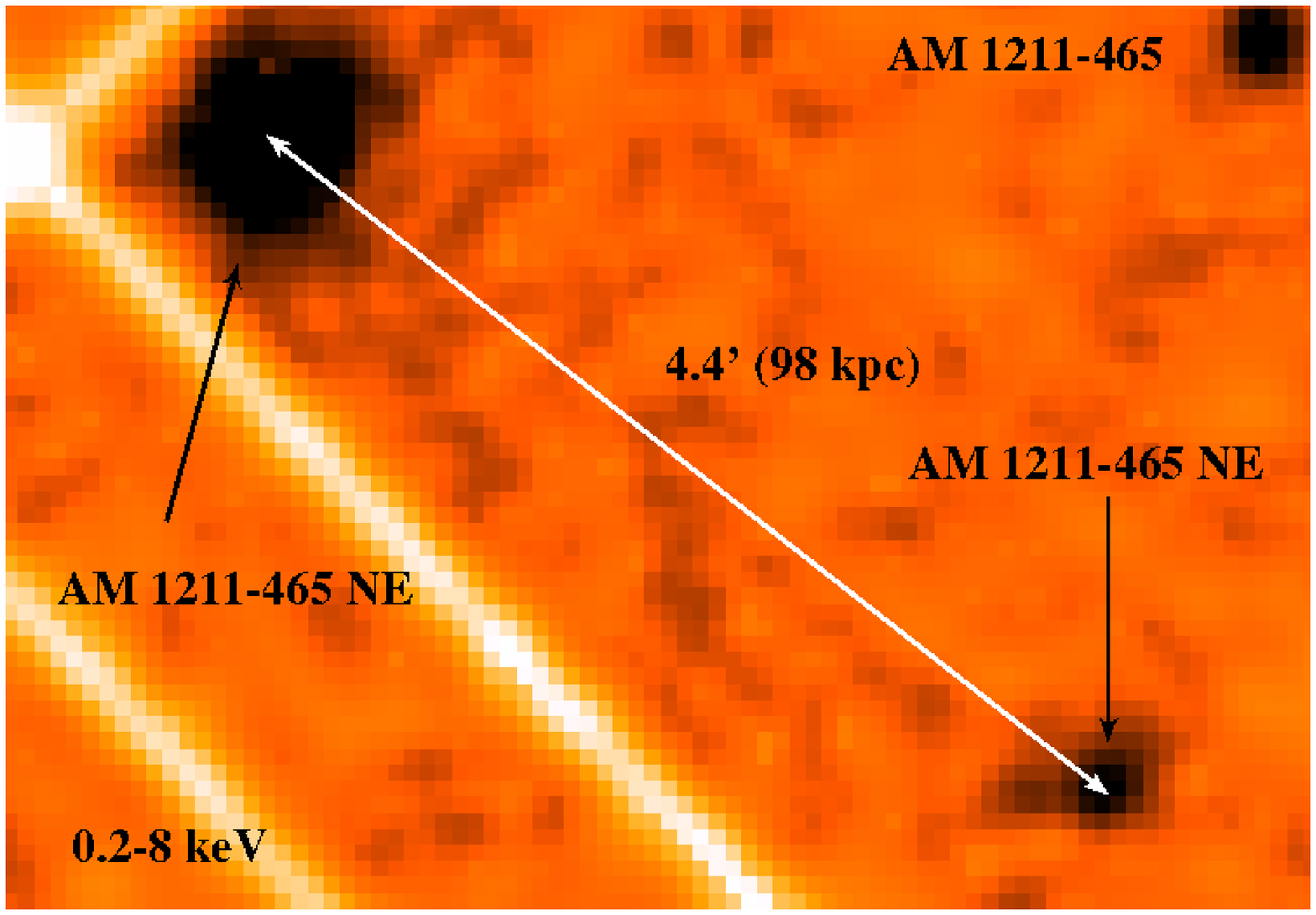}\includegraphics*[width=44mm,angle=0]{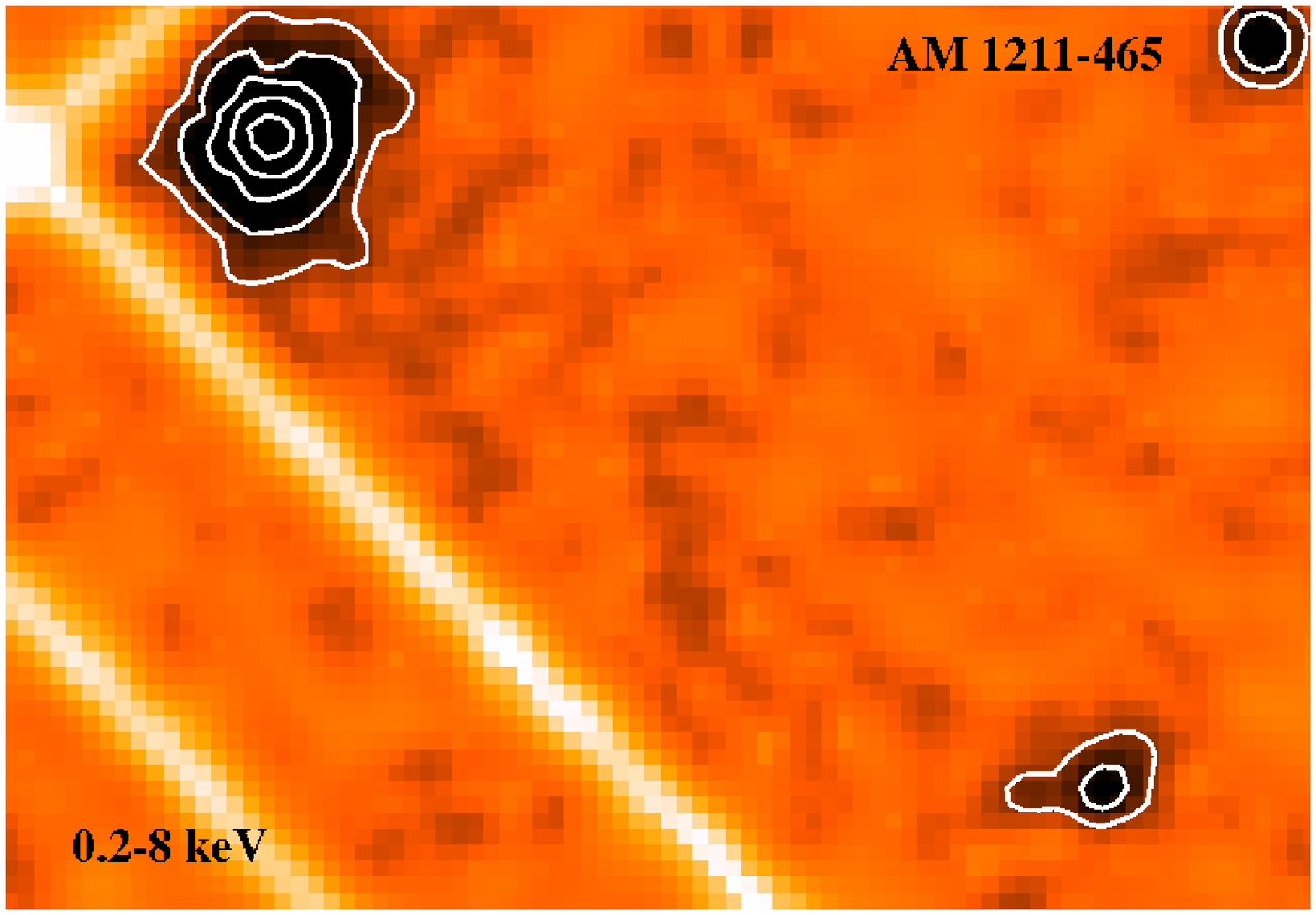}
\includegraphics*[width=44mm,angle=0]{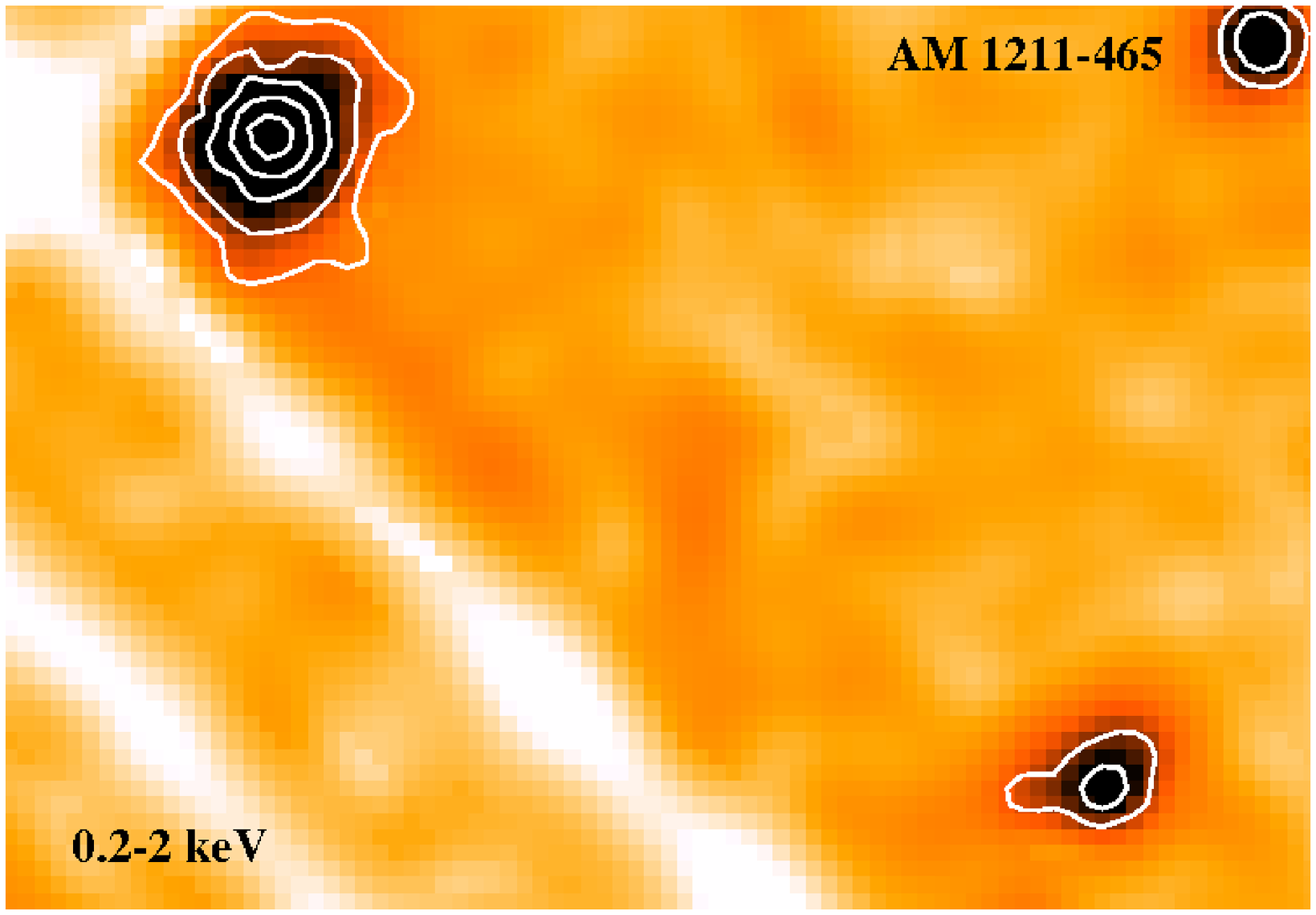}\includegraphics*[width=44mm,angle=0]{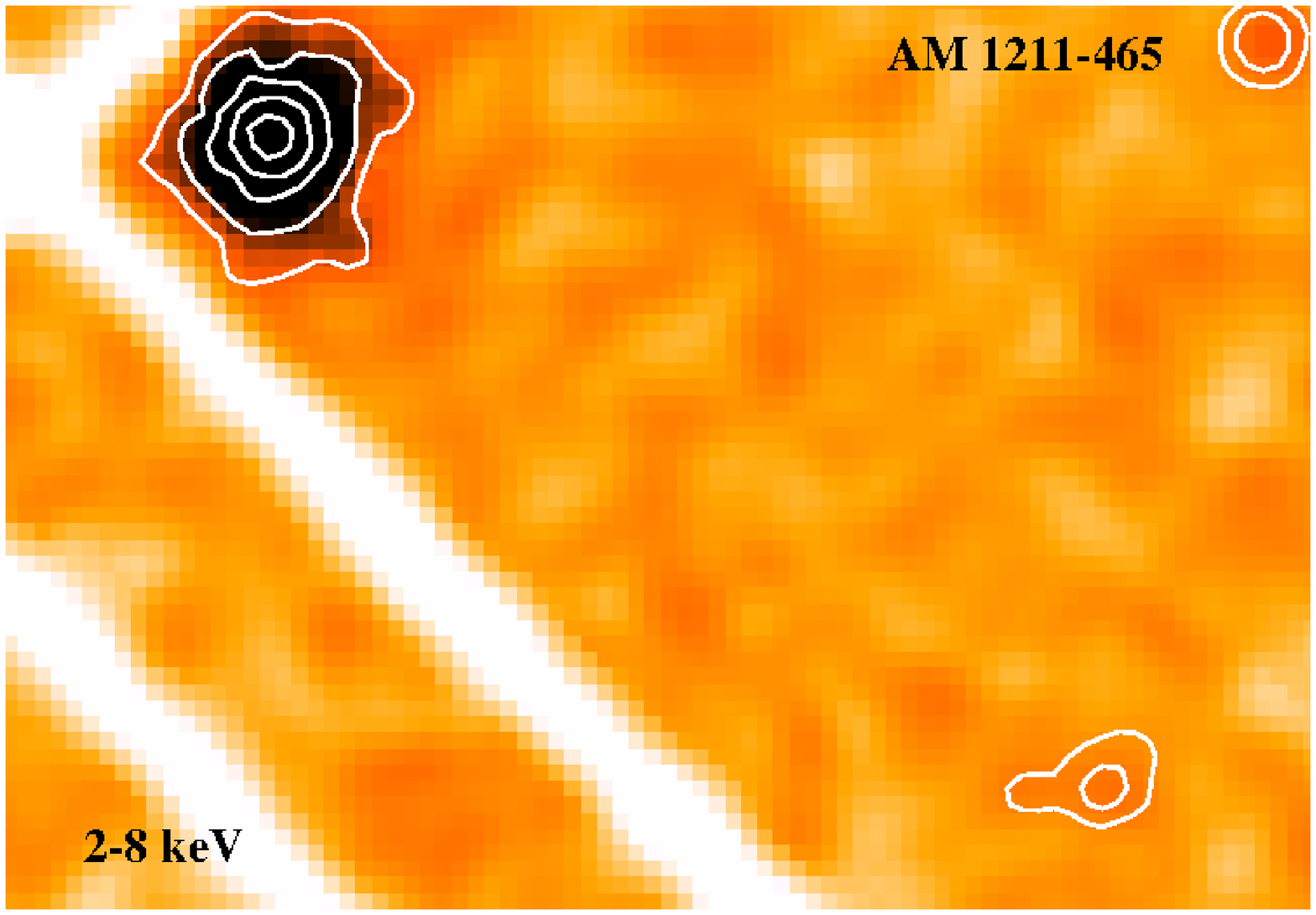}
\caption{Smoothed \pn\  images of AM~1211-465 on a logarithmic scale from 0 to 350 counts. From top  to bottom and
left  to right: (a) Location  and separation  distance of  the sources
superimposed  on  the smoothed 0.2-8~keV  image.  (b) 0.2-8~keV  surface
brightness contours superimposed  on the
smoothed  0.2-2~keV  image.  Contour levels are at 7, 10, 1, 20, 50, and 150 counts. (c) 0.2-8~keV  surface
brightness contours superimposed  on the
smoothed  2-8~keV  image.   (d) 0.2-8~keV  surface
brightness contours superimposed  on the
smoothed  0.2-8~keV  image. }\label{fig:am1211}
\end{figure}

\begin{figure}
\includegraphics*[width=44mm,angle=0]{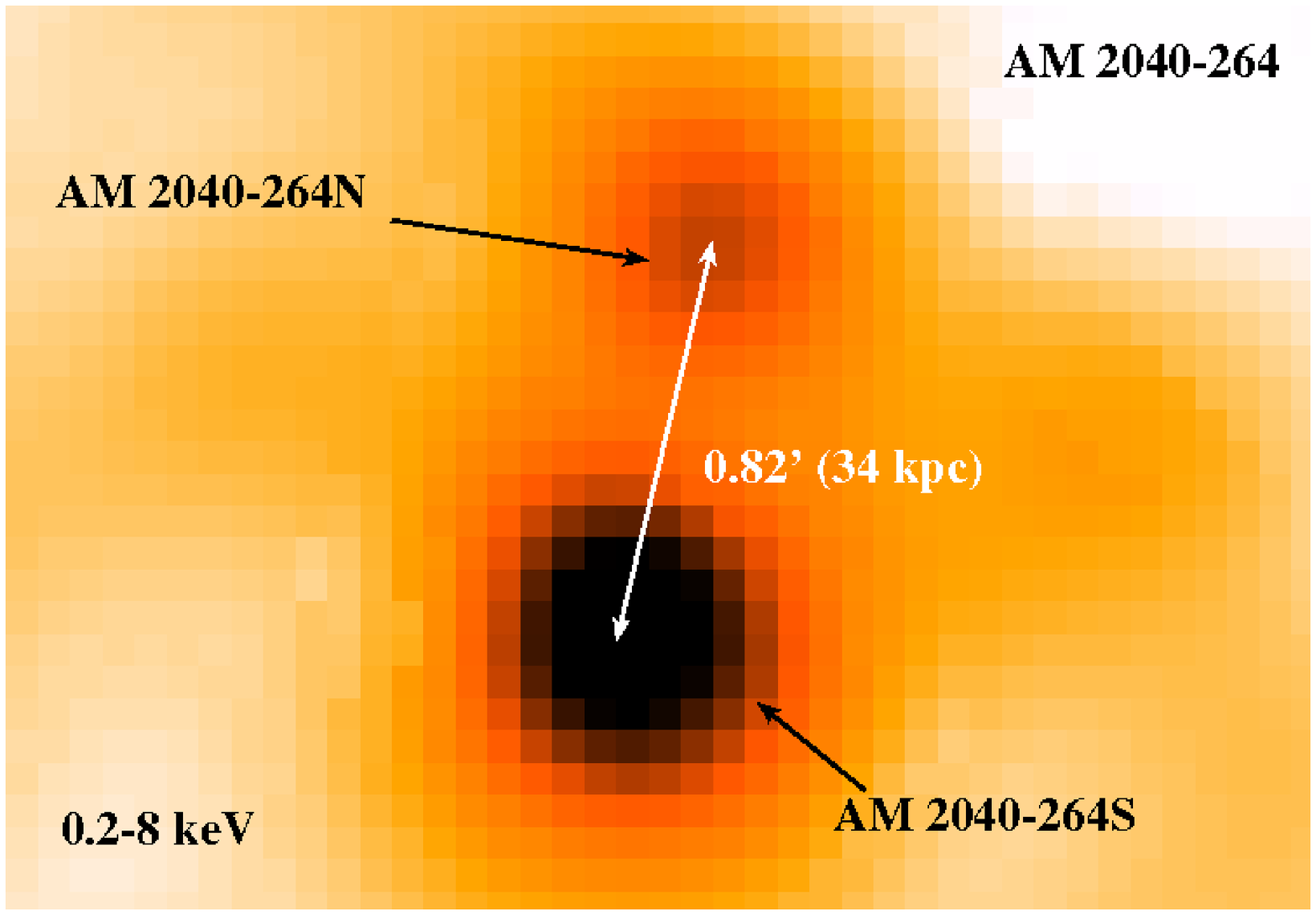}\includegraphics*[width=44mm,angle=0]{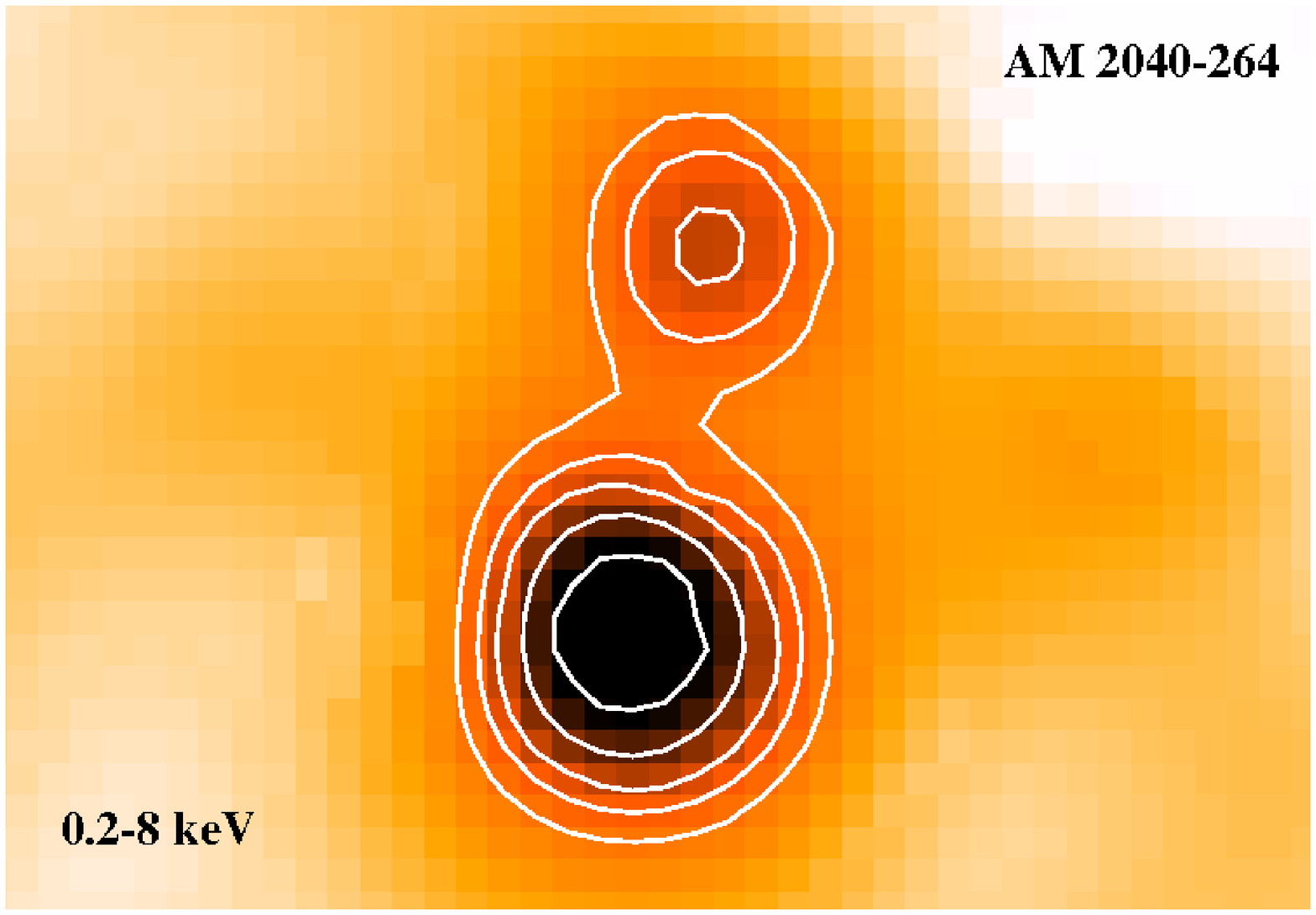}
\includegraphics*[width=44mm,angle=0]{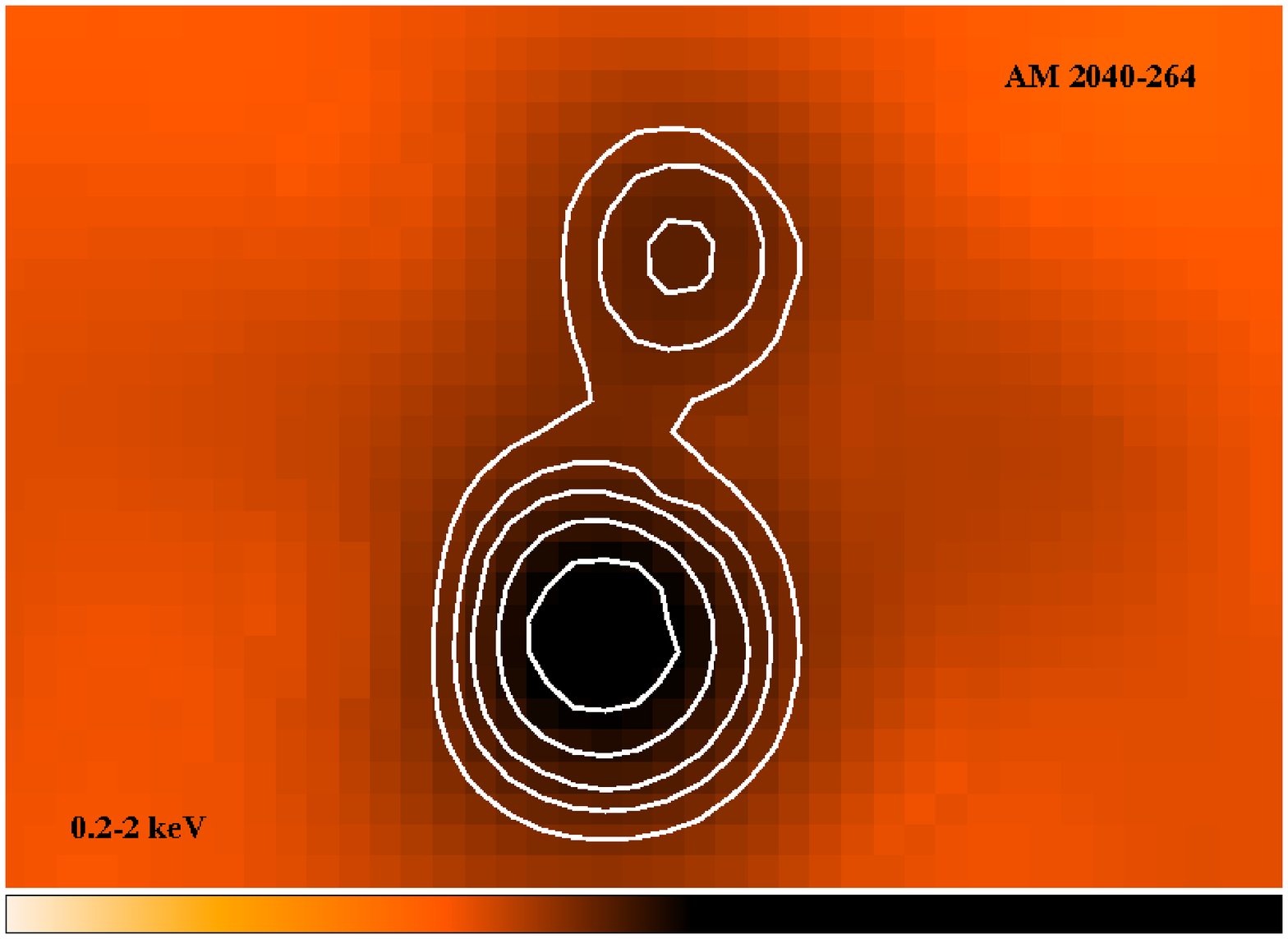}\includegraphics*[width=44mm,angle=0]{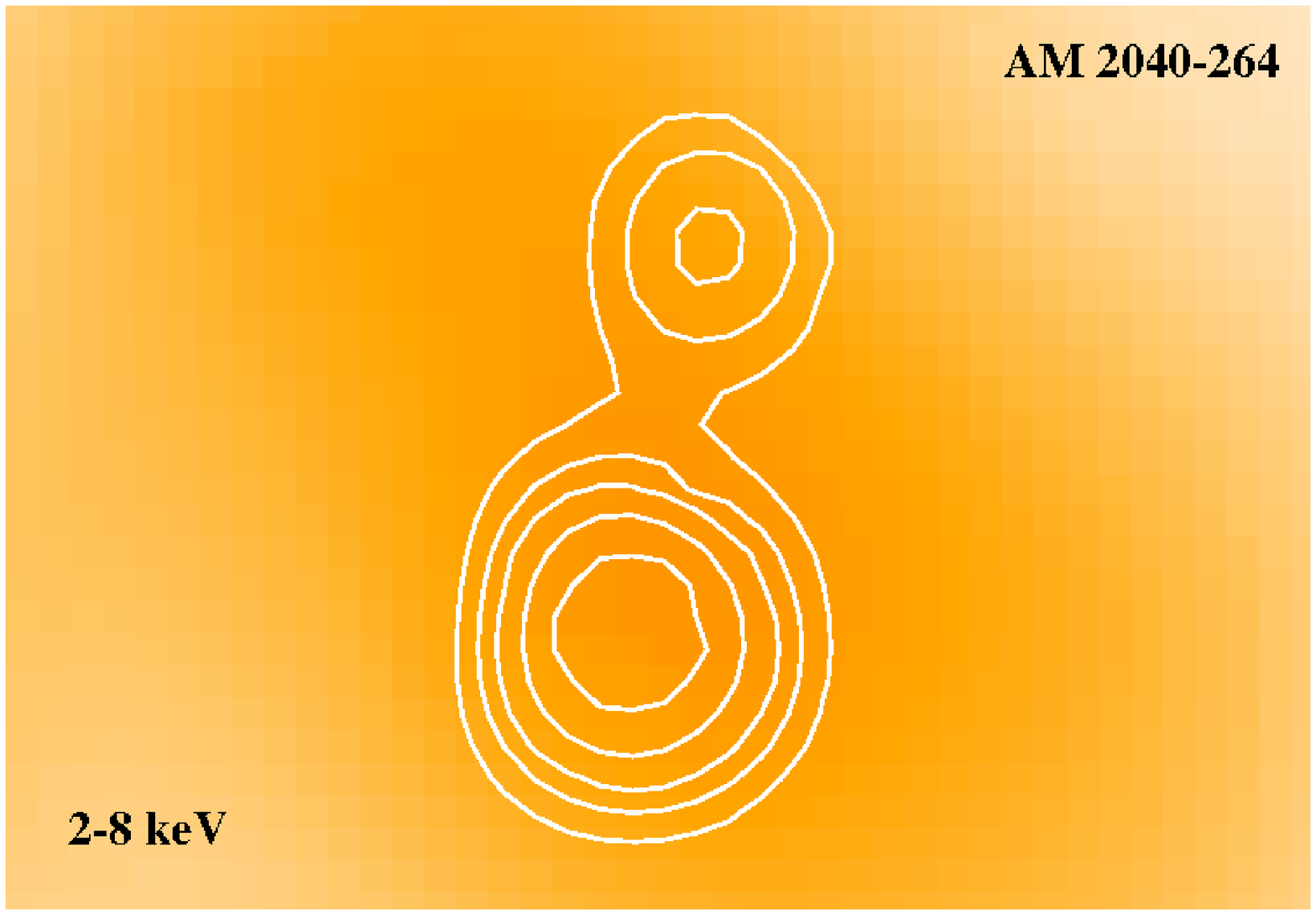}
\caption{Smoothed \pn\  images of AM~2040-674  on a logarithmic scale from 0 to 150 counts. From top  to bottom and
left  to right: (a) Location  and separation  distance of  the sources
superimposed  on  the smoothed 0.2-8~keV  image.  (b) 0.2-8~keV  surface
brightness contours superimposed  on the
smoothed  0.2-2~keV  image. Contour levels are at 0.0003, 0.003, 0.03, 0.3, and 3 counts. (c) 0.2-8~keV  surface
brightness contours superimposed  on the
smoothed  2-8~keV  image.   (d) 0.2-8~keV  surface
brightness contours superimposed  on the
smoothed  0.2-8~keV  image.}\label{fig:am2040}
\end{figure}

Smoothed  images of  the galaxy  pairs were  generated using  the {\it
asmooth} SAS  task applied  to the \pn\  images.  We applied  the {\it
adaptive} convolution technique,  designed for Poissonian images, with
S/N=10.          Figures~\ref{fig:am0707}a,         \ref{fig:am1211}a,
and~\ref{fig:am2040}a show the \pn\ 0.2-8~keV smoothed images of
each of  the targets.  All six  galaxies were clearly  detected in the
whole  energy band  and the  \xmm\  spatial resolution  allowed us  to
disentangle the  two members  in all pairs.   In the figures,  we have
named  each  pair  member   and  the  separation  distances  are  also
indicated.  In  Table~\ref{table:location}, we collate  the locations,
the separation distance, both in arcsec  and in kpc for all pairs, the
redshift,  and the  Galactic equivalent  hydrogen column  obtained from
Dickney  \&  Lockman (1990).  We  also  produced  images in  the  soft
(0.2-2~keV) and in the hard (2-8~keV) bands.

\begin{figure}
\includegraphics*[width=33mm,angle=90]{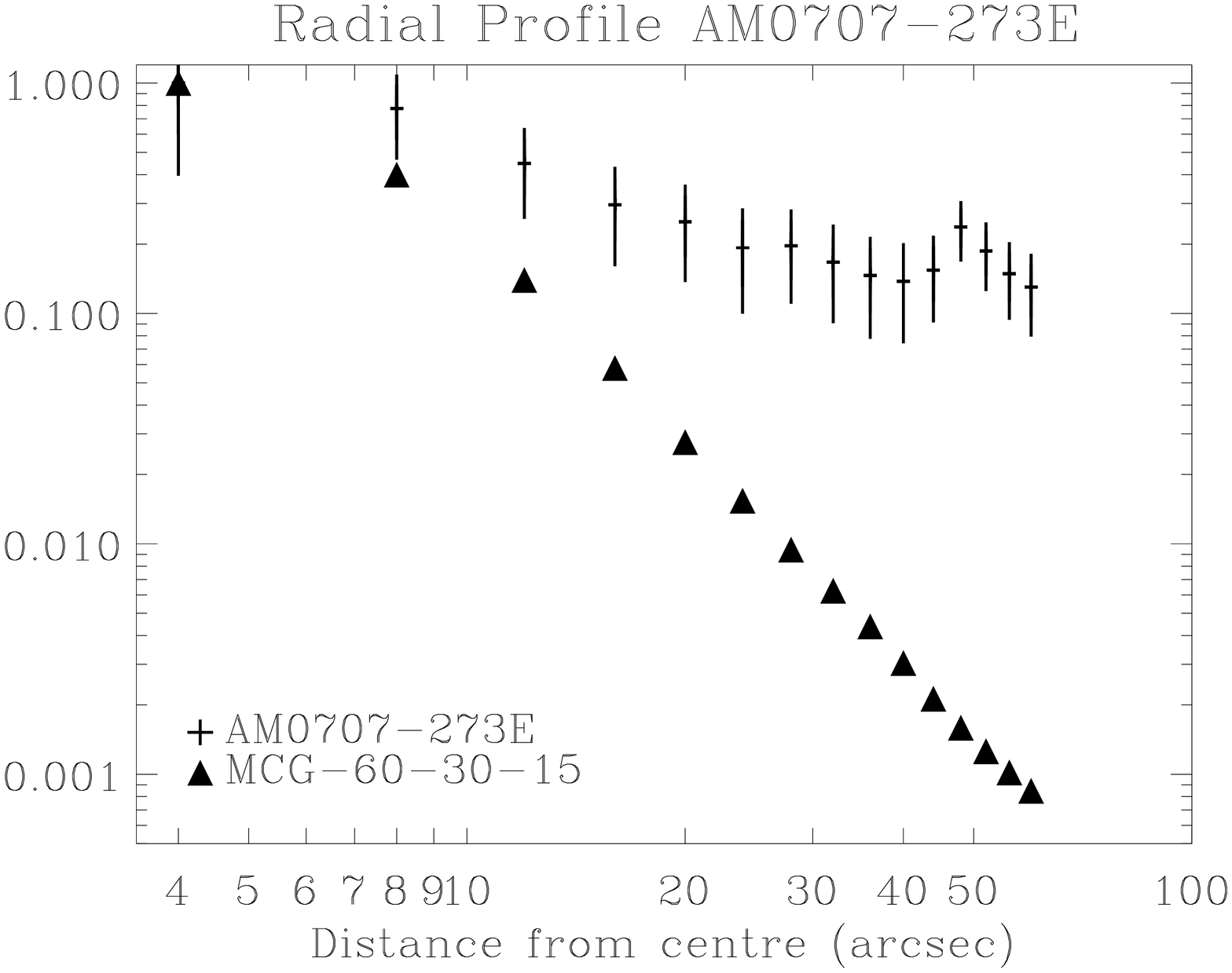}\includegraphics*[width=33mm,angle=90]{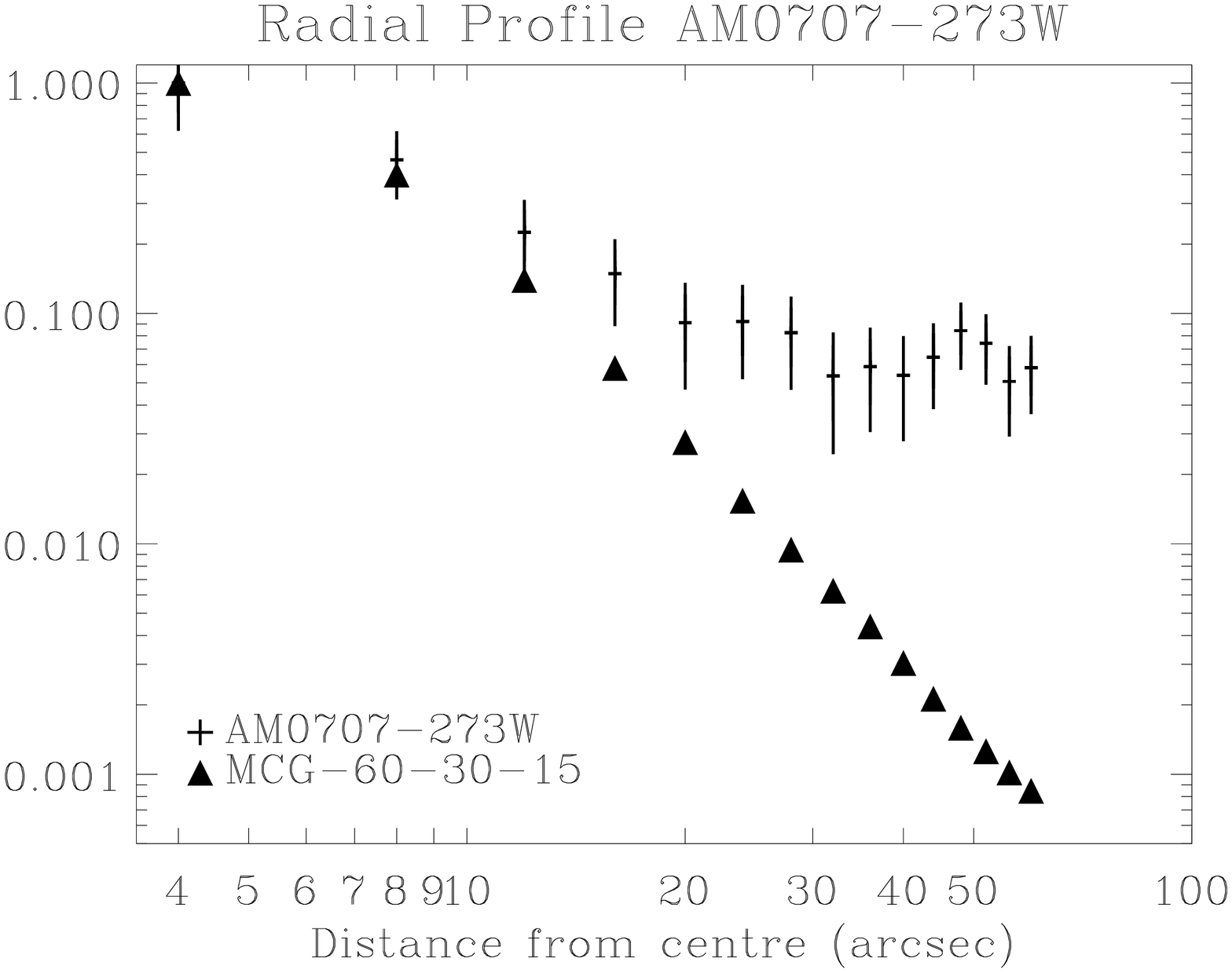}
\caption{Radial profile of AM~0707-273E (left panel), i.e. integrated
net counts normalized versus distance from the center of the galaxy,
and AM~0707-273W (right panel) in the 0.2-8~keV energy band.  Radial
profiles of the point-like source MCG-60-30-15 in the same energy bands are
also plotted for comparison. In both profiles, a local maximum of the
normalized net counts due to presence of the companion is observed
around 50\arcsec. }\label{fig:radial_profile_AM0707}
\end{figure}

Radial profiles  were produced  in  the 0.2-8~keV band   for both
AM0707- 273 members.  Despite of their proximity, which prevents us from reaching  firm conclusions,  both  sources seem  to  be extended  when
compared   with   the   point-like   profile   of   MCG-6-30-15   (see
Fig.~\ref{fig:radial_profile_AM0707}).   In  fact,  the  AM0707-273E
component presents  a more extended profile than  its companion.  This
extended nature of the emission gives hints of the merging process. As
seen in Fig.~\ref{fig:am0707}, both galaxies show intense nuclear
emission in the hard band.  

\begin{table*}[h!bt]
\caption{Locations  and separation  distances for  each of  the galaxy
pairs}\label{table:location}
\begin{tabular}{lllllll}
\hline
{\bf Target} & \multicolumn{2}{c}{\bf Coordinates J2000} & \multicolumn{2}{c}{\bf Separation} & {\bf Redshift} & {\bf n$_{\rm{H}}^{\rm {Gal}}$}\\
 & RA & Dec & Arcmin & kpc & (1) & $10^{20}$cm$^{-2}$ \\ \hline\hline
AM 0707-273E & 07 09 49\farcs6 & $-$27 34 35 & 0\farcm78 & 9   & $^a$ 0.00989$\pm(2\times10^{-5})$ & 1.7 \\
AM 0707-273W & 07 09 46\farcs6 & $-$27 34 09 &  & & $^b$ 0.00989$\pm(2\times10^{-5})$ \\\hline
AM 1211-465NE & 12 14 13\farcs0 & $-$47 13 45 & 4\farcm4 & 98  & $^c$ 0.01849$\pm(1.2\times10^{-4})$  & 8.4 \\
AM 1211-465SW & 12 14 52\farcs4 & $-$47 16 29  & & & $^d$ 0.01848$\pm(1.7\times10^{-4})$ \\\hline
AM 2040-674N & 20 45 19\farcs8 & $-$67 32 16 & 0\farcm82 & 34 &   $^e$ 0.03249$\pm(9\times10^{-5})$ & 4.3\\
AM 2040-674S & 20 45 21\farcs0 & $-$67 33 04  & & & $^e$ 0.03263$\pm(9\times10^{-5})$ \\\hline
\end{tabular}\\
{  \scriptsize Notes:  (1) Redshift values obtained from ($^a$) Theureau at al. (2005); ($^b$) Wong (2006); ($^c$) Strauss et al. (1992); ($^d$) Fouque et al. (1992); \\($^e$) Sekiguchi  \&
Wolstencroft (1992)} 
\end{table*}

\begin{figure}
\includegraphics*[width=33mm,angle=90]{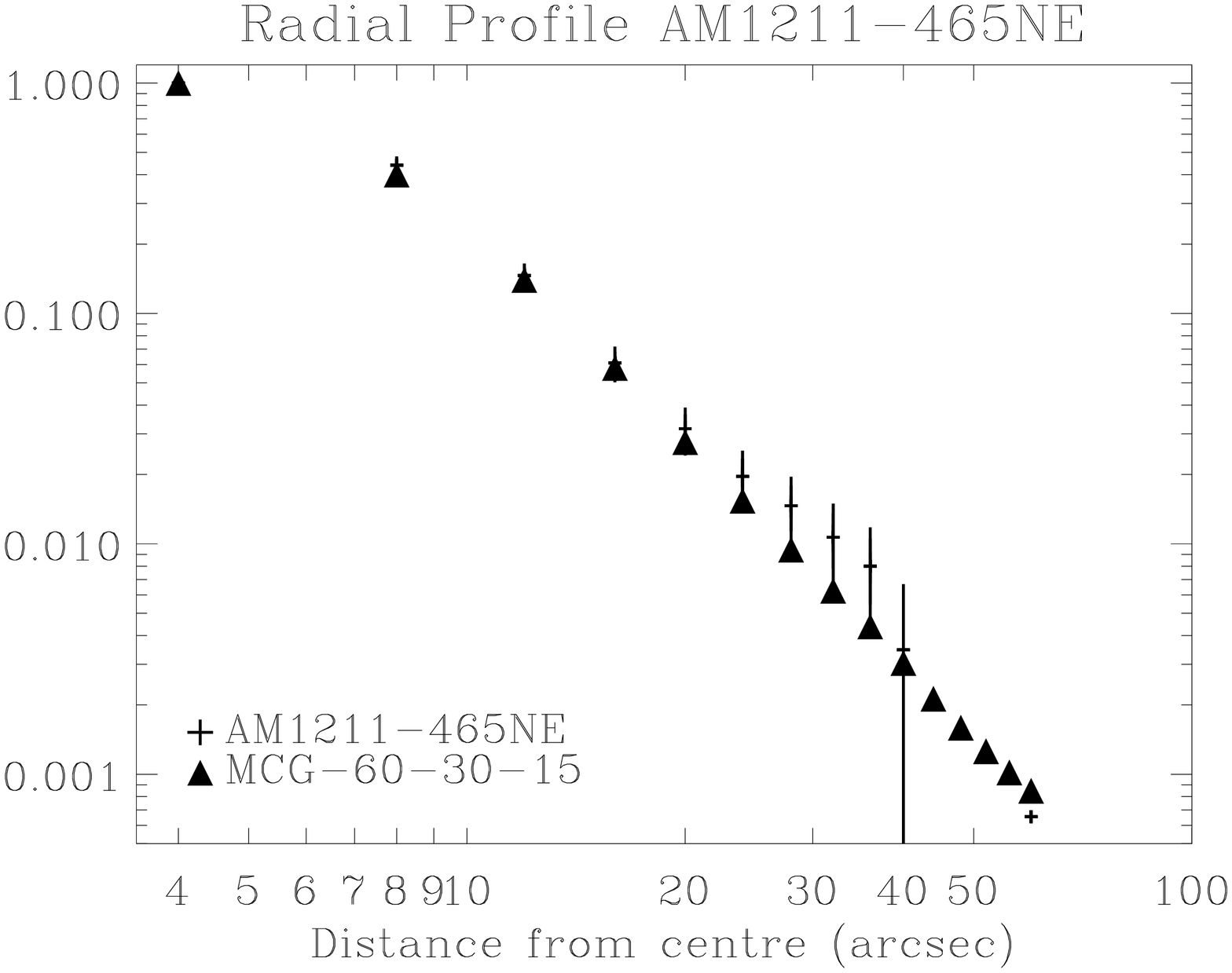}\includegraphics*[width=33mm,angle=90]{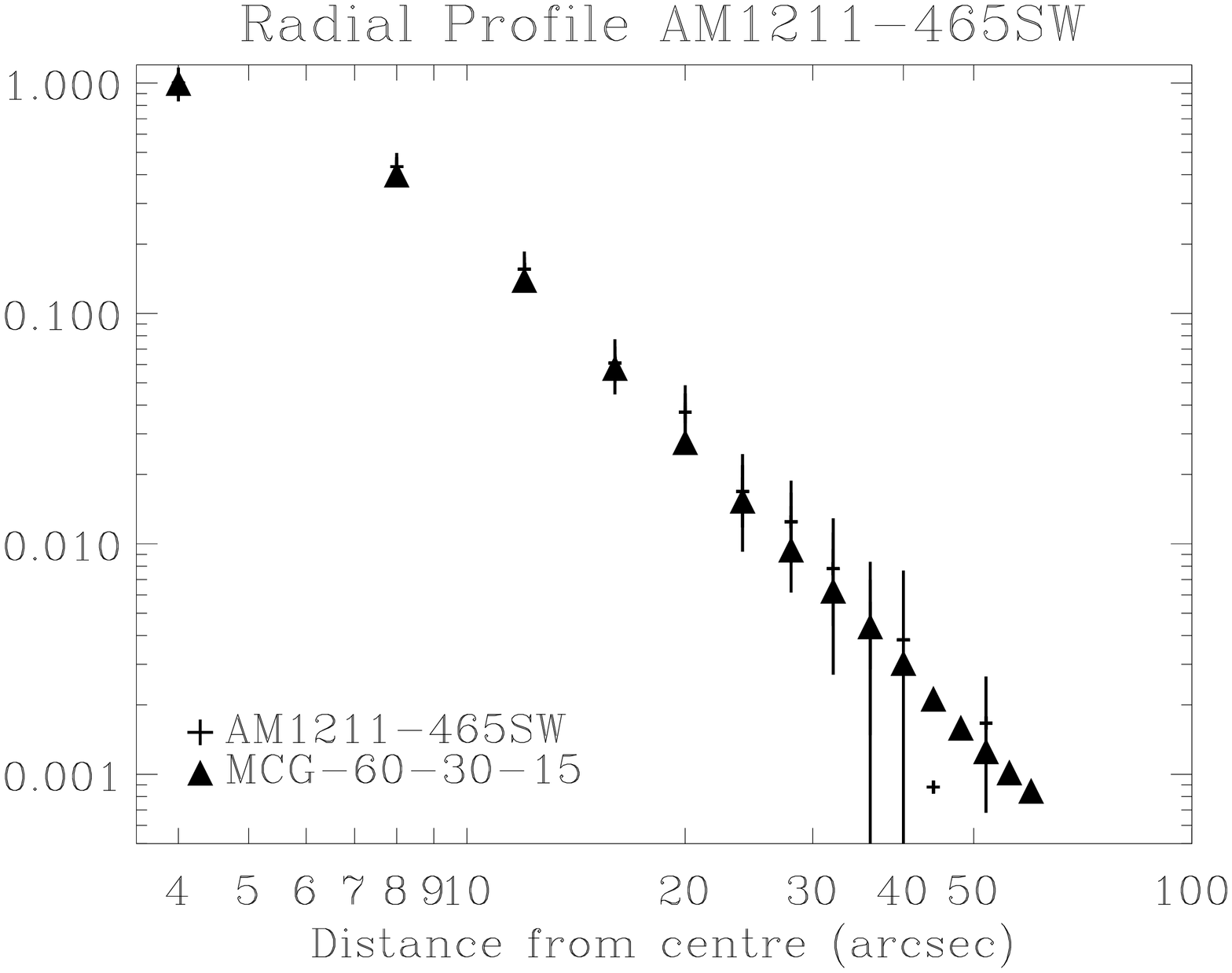}
\caption{Radial   profile   of    AM~1211-465NE   (left   panel)   and
AM~1211-465SW  (right  panel) in  the  0.2-8~keV energy  band.
Radial  profiles of  the point-like source  MCG-60-30-15 in  the same
energy         bands        are        also         plotted        for
comparison.}\label{fig:radial_profile_AM1211}
\end{figure}

In  the case  of AM1211-465,  the radial  profile indicates  that both
sources are point-like  in the broad 0.2-8~keV energy band (see
Fig.~\ref{fig:radial_profile_AM1211}).  Both  galaxies are detected in
the hard  energy band.  The  AM1211-465NE hard-band image  reveals the
presence  of a  very intense  X-ray source.   The  images  show a
region of diffuse emission  between the sources 10$\sigma$ above the
background, strengthening  the possibility of  an effective encounter
between the galaxies, even though  its apparent separation is high, 98
kpc.  Unfortunately, its weakness and the proximity of the edge
of the CCD to the sources do not allow  more accurate analysis of the
shape of this diffuse emission.

\begin{figure}
\includegraphics*[width=50mm,angle=90]{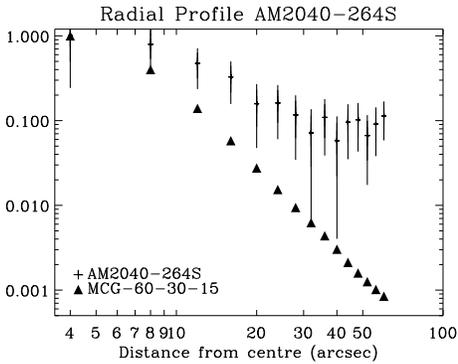}
\caption{Radial profile of AM~2040-674S in the 0.2-8~keV energy band.
Radial  profile of  the point-like  source MCG-60-30-15  in the  same
energy         bands        are        also         plotted        for
comparison.}\label{fig:radial_profile_AM2040}
\end{figure}

Finally, the two galaxies of AM2040-674 are visible in the soft energy
band but  AM2040-674N was undetectable  in the hard energy  band.  The
radial   profile    (Fig.~\ref{fig:radial_profile_AM2040})   is   not
conclusive, due to the  proximity of the sources ($\sim50\arcsec$) and
the low  number of counts  detected for the northern  member. However,
marginal hints  of extended  emission are shown  in the  AM2040- 674-S
radial profile.

\subsection{Spectral analysis} \label{sec:spectral_analysis}

We  extracted   the  \epic\  spectra  for  the   sources  with  enough
counts.The \mos1 and \mos2 spectra were  combined in order to increase the
signal-to-noise,  and the resulting  spectra were  analyzed together
with the  \pn\ one.   All \pn\ and  \mos1/2 spectra were  grouped such
that  each bin  contains at  least  20 counts  in order  to apply  the
modified $\chi^2$ minimization technique (Kendall et al.  1973) in the
spectral analysis.   For all cases,  the source extraction  region was
circular with  a radius chosen  to maximize the  signal-to-noise ratio
and to keep it from overlapping the extraction region of the companion
source.  The background extraction  regions were also circular, placed
in  the same  CCD  of  the targets  and  in a  field  free from  other
contamination sources.  Two of the pairs, AM~0707-273 and AM~2040-674,
show  separation between galaxies  on the  order of  50\arcsec. For
these  cases, the  extraction  radii are  20\arcsec.  The fraction  of
encircled energy  for the \pn\ detector within  30\arcsec\ is measured
to be higher than 80\% (\xmm\ User's Handbook V2.4); therefore, we
did not  expect a significant contamination from  the companion source
in  the spectrum.   Only for  one of  the galaxies,  AM~2040-674N, we did not
detect enough  counts to perform a  spectral analysis.  The
spectral analysis was performed using XSPECv.12.2.0 (Arnaud 1996). The
quoted errors for  the fit parameters referred to  the 90\% confidence
level  (i.e.  $\Delta\chi^2=2.71$,  Avni 1976).   We assumed  a Hubble
constant    of    70~kms$^{-1}$   and    a    flat   cosmology    with
$\Omega_M,\Omega_{\Delta}=(0.3,0.7)$ (Bennett et al. 2003).

\begin{table*}[htb]

\caption{Results of the spectral analysis of the five galaxies studied.}\label{tab:models}

\begin{tabular}{lllllllllll}
\hline 

 {\bf Target} & {\bf Model} & n$_{\rm {H}}$ & $\Gamma$/kT &    n$_{\rm {H}}$ &kT & EW(Fe line) & Significance & Goodness \\
 & (1) & \\
& & 10$^{21}$cm$^{-2}$ & &  10$^{21}$cm$^{-2}$ & keV & eV\\

\hline\hline

AM 0707-273-E & A & $1.1\pm0.6$ & $1.9^{+0.3}_{-0.2}$ & - & - &- & -& 33 for 22 dof \\
 & B & $\le0.3$ & $6^{+4}_{-2}$ &  - & - & - & -  &35 for 22 dof \\
 & {\bf C} & $\sim0$ & $1.2\pm0.2$ &   $10.4^{+1.1}_{-1.9}$  &$0.143^{+0.07}_{-0.002}$ & $<2400$ & 89\%  &  {\bf 19 for 20 dof} \\ 
 & D & $\sim0$ & $1.2^{+0.2}_{-0.7}$ &  $9^{+20}_{-8}$& $0.16^{+0.5}_{-0.05}$ &- &- & 21 for 20 dof \\
 & E & $<1.2$ & $\ge7$ &  $\sim0$  & $0.85^{+0.2}_{-0.16}$ & - & - & 20 for 20 dof\\
AM 0707-273-W & A & $2.8^{+0.7}_{-0.8}$ & $3.1^{+0.4}_{-0.3}$ &- &   -&- &- & 56 for 30 \\
& B & 0.15 & 3 &  -  & - & - & -  & 87 for 30 dof \\
 & {\bf C} & $1.3^{+1.9}_{-1.1}$ & $2.2^{+0.8}_{-0.5}$ &   5$^{+4}_{-3}$ & $0.43^{+0.15}_{-0.2}$ &$<9000$  & 80\% & {\bf 40 for 27 dof} \\
 &  D & $2.6^{+4}_{-1.5}$ & $2.7^{+0.6}_{-0.3}$ & $3.0\pm0.9$  & $0.45^{+0.17}_{-0.10}$ & - & - & 46 for 27 dof \\ 
& E & $0.7^{+1.5}_{-0.4}$ & $3.6^{+1.2}_{-1.0}$ &  $\sim0$  & $0.59^{+0.08}_{-0.12}$ &- &- & 49 for 27 dof \\\hline

AM 1211-465-NE & A & 12 & 1.3 &- & -& -&- &531 for 214 dof \\
 & B & 11 & $>80$ &  -  & - & - & -  & 547 for 214 dof \\
 & {\bf C} & $22\pm2$  & $1.63\pm0.10$ & $1.2^{+1.3}_{-0.4}$  & $0.52^{+0.11}_{-0.19}$ &$110\pm70$ & 98.8\%& {\bf 228 for 209 dof} \\
& D & $22^{+5}_{-3}$ & $1.64^{+0.12}_{-0.13}$ &  $4^{+3}_{-2}$  & $0.25^{+0.2}_{-0.14}$ &  - & - & 241 for 211 dof\\
 & E & $13.1^{+1.4}_{-0.5}$ & $\ge65$ & $\sim0$ & $0.40^{+0.10}_{-0.06}$ &- & -& 294 for 212 dof\\

AM 1211-465-SW & A & $\sim0$ & $1.7\pm0.2$ & - & - & - & - & 22 for 21 dof \\ 
 & B & $\sim0$ & $8^{+16}_{-3}$  & -  & - & - & -  & 26 for 21 dof \\
 &  {\bf C} & $\sim0$ & $1.0\pm0.5$ & $8.0^{+7}_{-1.4}$ & $<0.2$ & $<3000$ & 72\% & {\bf 14 for 18 dof} \\
&  D & $\sim0$ & $0.3^{+0.5}_{-0.3}$ & $<7$ & $0.6^{+1.6}_{-0.4}$ & - & - & 14 for 18 dof\\
 & E & 8 & $>12$ & $7.9^{+1.6}_{-1.3}$ & $<0.3$ & -  & - & 14 for 18 dof \\\hline


AM 2040-674-S & A & $1.1^{+0.8}_{-0.6}$ &$3.2^{+0.6}_{-0.4}$  & - & - & - & - & 22 for 12 dof\\
 & B &$\sim0$  & 3 &  -  & - & - & -  & 64 for 13 dof \\
& {\bf C }& $\sim0$ & $1.9^{+0.4}_{-0.2}$ & $\sim0$ & $0.33^{+0.18}_{-0.06}$  & $<4\times10^{6}$ & 50\% & {\bf 9 for 11 dof}\\ 
&  D &  $\sim0$& 1.4 & $\sim0$  & 0.5 & -  & -  & 22 for 11 dof\\
& E & $\sim0$ & $7^{+40}_{-4}$ & $\sim0$ & $0.32^{+6}_{-0.05}$ & - & - & 10 for 11 dof\\
\hline
\end{tabular}

{\scriptsize Notes: (1) Model A: wabs*pwlw; Model B: wabs*mekal; Model C: wabs*pwlw+wabs*mekal; Model D: wabs*pwlw+wabs*bremsstrahlung; Model E: wabs*(mekal+mekal). {\it wabs} represents photoelectric absorption (Morrison \& McCammon, 1983) and {\it pwlw} a power law emission model. The best fit model is marked in boldface.}
\end{table*}

\begin{table*}[htb]
\caption{Absorbed fluxes and unabsorbed luminosities for the six galaxies.}\label{tab:lum}
\begin{tabular}{lllll}
\hline
{\bf Target} & {\bf Flux (0.5-2 keV)} & {\bf Flux (2-10 keV)} & {\bf Luminosity (0.5-2 keV)} & {\bf Luminosity (2-10 keV)} \\
 & 10$^{-14}$ erg cm $^{-2}$s$^{-1}$ & 10$^{-14}$ erg cm $^{-2}$s$^{-1}$ & 10$^{41}$ erg s$^{-1}$ & 10$^{41}$ erg s$^{-1}$ \\
\hline\hline
AM 0707-273-E & $2.89^{+0.11}_{-1.6}$ & $6.9^{+1.0}_{-2}$ & $33.6^{+1.3}_{-20}$ & $0.15^{+0.02}_{-0.04}$  \\
 \,\,\, {\it pwlw} & 1.93(70\%) & 6.9 ($\sim$100\%)& 0.05 (-) & 0.15 ($\sim$100\%)\\
\,\,\, {\it mekal}& 0.96 (30\%)& 8$\times10^{-4}$ (-) & 33.6 ($\sim$100\%) & 3$\times10^{-4}$ (-)\\
AM 0707-273-W & $4.3^{+0.2}_{-2}$ &  $3.7^{+0.6}_{-1.5}$ &  $0.375^{+0.017}_{-0.17}$ & $0.082^{+0.013}_{-0.03}$ \\
\,\,\, {\it pwlw} & 2.9 (70\%) & 3.6 (99\%) & 0.17 (45\%) & 0.08 (98\%)\\
\,\,\, {\it mekal}&  1.4 (30\%) & 0.05(1\%) & 0.205 (55\%) & 1.3$\times10^{-3}$ (2\%)\\ \hline
AM 1211-465-NE & $15.6^{+0.7}_{-3}$ & $214^{+13}_{-17}$ & $10.6^{+0.5}_{-2}$ & $19.4^{+1.1}_{-1.5}$ \\
 \,\,\, {\it pwlw} & 11 (70\%)& 212 ($\sim$100\%)  & 9.8 (90\%) & 19.4 ($\sim$100\%) \\
\,\,\, {\it Raymond-Smith}& 4.6 (30\%)  & 3$\times10^{-3}$ (-) & 0.8($\sim$10\%)  & 0.02 (-) \\
AM 1211-465-SW & $3.4^{+0.5}_{-2}$ & $12^{+2}_{-4}$ & $65^{+10}_{-40}$ & $0.90^{+0.15}_{-0.3}$ \\ 
 \,\,\, {\it pwlw} & 2.0 (60\%)& 12 ($\sim$100\%)  & 0.2 (-) & 0.90 ($\sim$100\%) \\
\,\,\, {\it mekal}& 1.4 (40\%)  & 5$\times10^{-4}$ (-) & 65($\sim$100\%)  & 7$\times10^{-5}$ (-) \\ \hline
AM 2040-674-N$^{\dagger}$ & 0.9 & 1.8 & 0.28 & 0.43 \\
AM 2040-674-S & $1.9^{+0.3}_{-0.2}$ & $1.7^{+0.3}_{-0.7}$ & $0.83^{+0.13}_{-0.09}$ & $0.41^{+0.07}_{-0.17}$ \\ 
 \,\,\, {\it pwlw} & 1.2 (60\%) & 1.7 ($\sim$100\%) & 0.51 (60\%)& 0.41 ($\sim$100\%)\\
\,\,\, {\it Raymond-Smith} & 0.7 (40\%)& 3$\times10^{-3}$ (-)&  0.32 (40\%) & 9$\times10^{-4}$ (-)\\ \hline
\end{tabular}

{\scriptsize Notes: $\dagger$ Upper limits fluxes and luminosities calculated using the net \pn\ count rate measured, i.e. $7.1\times10^{-3}$ c/s, and assuming an absorbed \\
n$_{\rm {H}}=$1$\times10^{21}$~cm$^{-2}$ $\Gamma=1.8$ power law emission model}.
\end{table*}

To  elucidate the  nature of  the sources,  several spectral
models      have      been       tested      for      each      source
spectrum. Table~\ref{tab:models} summarizes  all the tested models, the
values  of their  parameters, and  the goodness  of the  fits  for each
source.   Single-component models  (power law  or thermal  {\it mekal}
emission) did  not provide  a satisfactory fit  for any of  the spectra
analyzed, excluding  the AM1211-465SW spectrum,  satisfactorily fitted
with  a  single   power  law  (see  below).   We   also  applied
two-component models  to all  spectra: a power  law and  two different
thermal models ({\it mekal} and {\it bremsstrahlung}).   The power-law component accounts  for either a possible AGN and,  in the case of
starburst origin  of the emission,  for the joint contribution  of the
emission  from X-ray  binaries and  a less  significant  contribution of
Compton radiation  that originated in  the interaction between  FIR photons
with  high-energy  cosmic  rays  from SN  ejecta.   The thermal  model
accounts for the joint  contribution of the compact stellar-like X-ray
sources and the  warm and ionized gas in superwinds.    The values of
the best-fit parameters span  wide ranges.  In  particular, the
power-law index varies from object  to object in the 1.0 to 2.2 range,
when only the best-fit model is taken into account.  The flatter power
laws are  found for both  galaxies in AM1211-465 and  for AM0707-273E.
In the  two remaining cases,  i.e.  AM0707- 273W and  AM2040-674N, the
index of the power law is compatible with values $\lesssim$ 2.

The  excess observed  in the  soft energy  band can  be satisfactorily
fitted  with  thermal emission  models.  Similar  $\chi^2$ values  are
obtained   for  the   two  tested   models,  {\it   mekal}   and  {\it
bremsstrahlung} for  both members of AM1211-465  and AM0707-273E.  The
$\chi^2$ values  for the {\it  mekal} and {\it  bremsstrahlung} models
differ slightly   for   AM0707-273W  and  AM2040-674S.   Although  the
temperatures  of the thermal  components that largely  vary from  object to
object,  the values  are  compatible  with the  ones  found in  nearby
interacting systems,  which are well-fitted  with temperatures ranging
from $\sim$0.15 to $\sim$0.9 keV  (Jenkins et al.  2004 and references
therein).   The soft  X-ray  emission of  merging  systems is  usually
well-fitted with multi-temperature  thermal components (Jenkins et al.
2004, and references therein).   Nonetheless, the addition of a second
thermal  component   to  account  for   the  soft  emission   did not
significantly improve any of the spectral fits.

The abundances  were initially fixed  to the solar value.  In general,
when they were left free to  vary, no bounded values were found.  Only
in the  case of AM1211-465NE,  does a different metallicity from  the solar
one significantly improve  the fit, Z=0.030$^{+0.03}_{-0.018}$, which makes  this  model  physically  indistinguishable  from  the  {\it
bremsstrahlung}  model.   The fitting  analysis  reveals that  similar
values  of the  \chired\  are found  for  the two  different types  of
thermal models applied.

A neutral  absorbing gas-material  component was tested  to separately
affect  each  of  the  spectral  components  (power  law  and  thermal
emission).  In some cases, we  found values of the N$_H$ close to
zero, so we removed the absorbing component from the testing
model. Only in the case  of AM1211-465NE, was a large equivalent hydrogen
column,  n$_{\rm{H}}$=2.2$\pm0.2\times10^{22}$   cm$^{-2}$, 
measured.     A     large    absorption,    n$_{\rm{H}}$=1.04$^{+0.11}
_{-0.19}\times10^{22}$  cm$^{-2}$,  affecting  the  thermal  component
was measured in AM0707-273E. For the rest  of the objects, in the cases in
which an absorbing component is  required, the values found are on the
order of  10$^{21}$ cm$^{-2}$.  The absence of  absorbing material for
AM2040-674S could indicate that the merging process is not effectively
obscuring the systems. 

We  also systematically tested the presence  of neutral K$\alpha$
iron line. This  feature is expected to be  present in highly-absorbed
AGN,  and  therefore its  occurrence  would  provide  evidence of  the
presence of Seyfert nucleus.  A  narrow Gaussian line fixed to 6.4~keV
significantly improves the $\chi^2$ only in the AM1211-465~NE spectrum
with a  confidence level of $\gtrsim$98.8\%, according  to the F-test.
A   lower  confidence   level   value,  but   still  relatively   high
($\sim$90\%), was found  for the galaxy AM0707- 273E.  The equivalent width (EW) measured was $110\pm70$~eV, fully in agreement with the mean EW of a sample of quasars observed with \xmm\ (Jim\'enez-Bail\'on et al. 2005). Upper limits to
the EW are calculated in the rest of the objects.

Finally, we  tested a double thermal {\it mekal} emission model to
explain the whole energy spectrum.  The one with the lower temperature
associated  with the  heated gas  related  to the  superwinds and  the
higher temperature  one associated with the  collisionally excited gas
created  by stellar  winds  and  SNe.  
  
The double  thermal model is not the  best model in any  of the cases;
however,  the  results   are  still  statically  acceptable.   Bounded
measurements  of the temperature  can be  derived for  AM0707-273W and
AM2040-674S,  (3.6$^{+1.2}   _{-1.0}$  keV  and   7$^{+40}_{-4}$  keV,
respectively).   These  values are  consistent  with  those found  for
starburst galaxies (Persic \& Rephaeli,  2002; Cappi et al. 1999). For
the  rest  of  the  objects,  i.e.   AM0707-  273E,  AM1211-465NE  and
AM1211-465SW, only lower limits for the temperature were derived.

Fluxes and luminosities in soft and hard energy bands derived from the
best-fit models are  given in Table~\ref{tab:lum}. The luminosities of
the   galaxies   in   the    hard   band   (2-10   keV)   range   from
$\sim$8$\times10^{39}$    to   $\sim$2$\times10^{42}$    ergs/s.    In
comparison  with several  samples  of LINER  and  HII galaxies,  these
luminosities are  in general  above the mean  values. Ho et  al.  (2001)
calculated  using \chandra\  a mean  L$_{\rm {2-10\,keV}}$  = 2$\times
10^{38}$  erg/s  for  a  sample  of LINER  and  composite  LINER/HII
galaxies.  Recently Gonz\'alez-Mart\'{\i}n  et al. (2006) calculated a
mean L$_{\rm {2-10\,keV}}$ of 3.8$\times10^{39}$ erg/s for a sample of
51   optically   selected   LINER,   covering   the   range   between
$1.2\times10^{38}-1.5\times10^{42}$~erg/s.

The  X-ray light curves in  different energy  bands, i.e.  0.3-10~keV,
0.5-2~keV,  and 2-10~keV, were  produced for the  five objects  with a
sufficient number  of detected counts. No signs  of variability within
the observation were observed for any of the objects.

\begin{figure*}
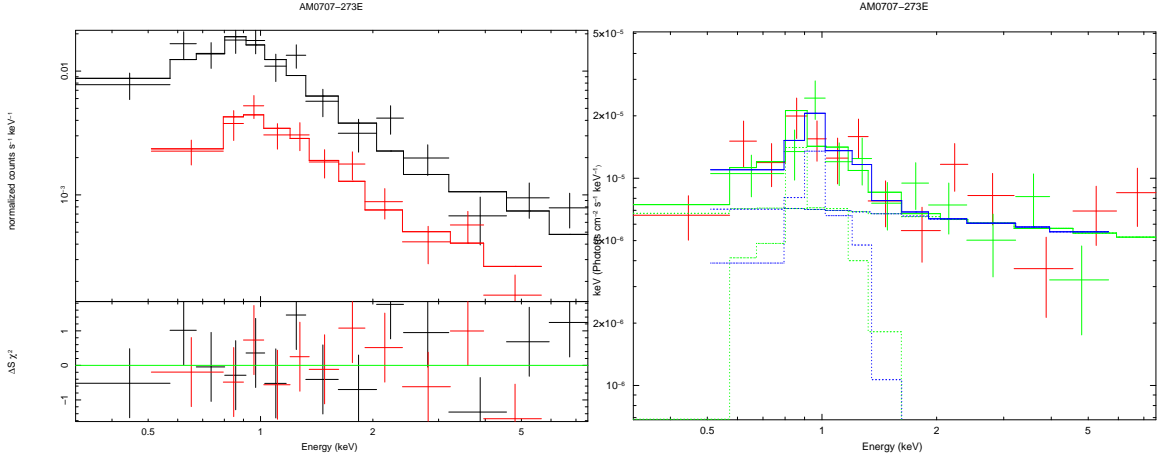

\includegraphics*[width=60mm,angle=-90]{am0707E_spec.ps}\includegraphics*[width=60mm,angle=-90]{am0707E_euf.ps}\\

\caption{Observed spectrum, best fit-model, and residuals of {\it AM 0707-273-E} on the left and the unfolded spectrum and the model in Ef(E) on the right. }

\label{fig:am0707E_spectra}
\end{figure*}

\begin{figure*}
\includegraphics*[width=60mm,angle=-90]{am0707W_spec.ps}\includegraphics*[width=60mm,angle=-90]{am0707W_euf.ps}

\caption{Observed spectrum, best-fit model, and residuals of {\it AM 0707-273-W} on the left and the unfolded spectrum and the model in Ef(E) on the right. }

\label{fig:am0707W_spectra}
\end{figure*}


\begin{figure*}
\includegraphics*[width=60mm,angle=-90]{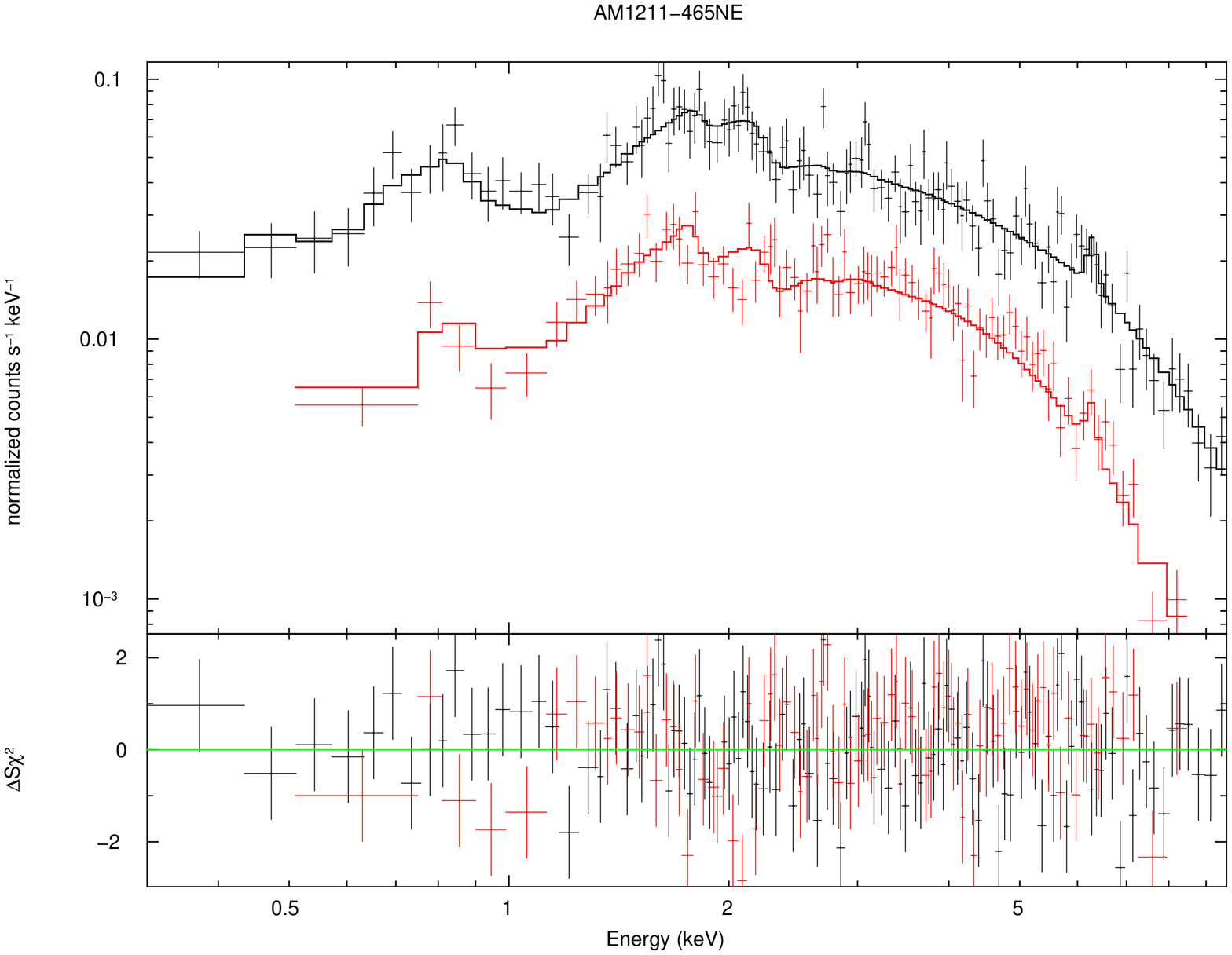}\includegraphics*[width=60mm,angle=-90]{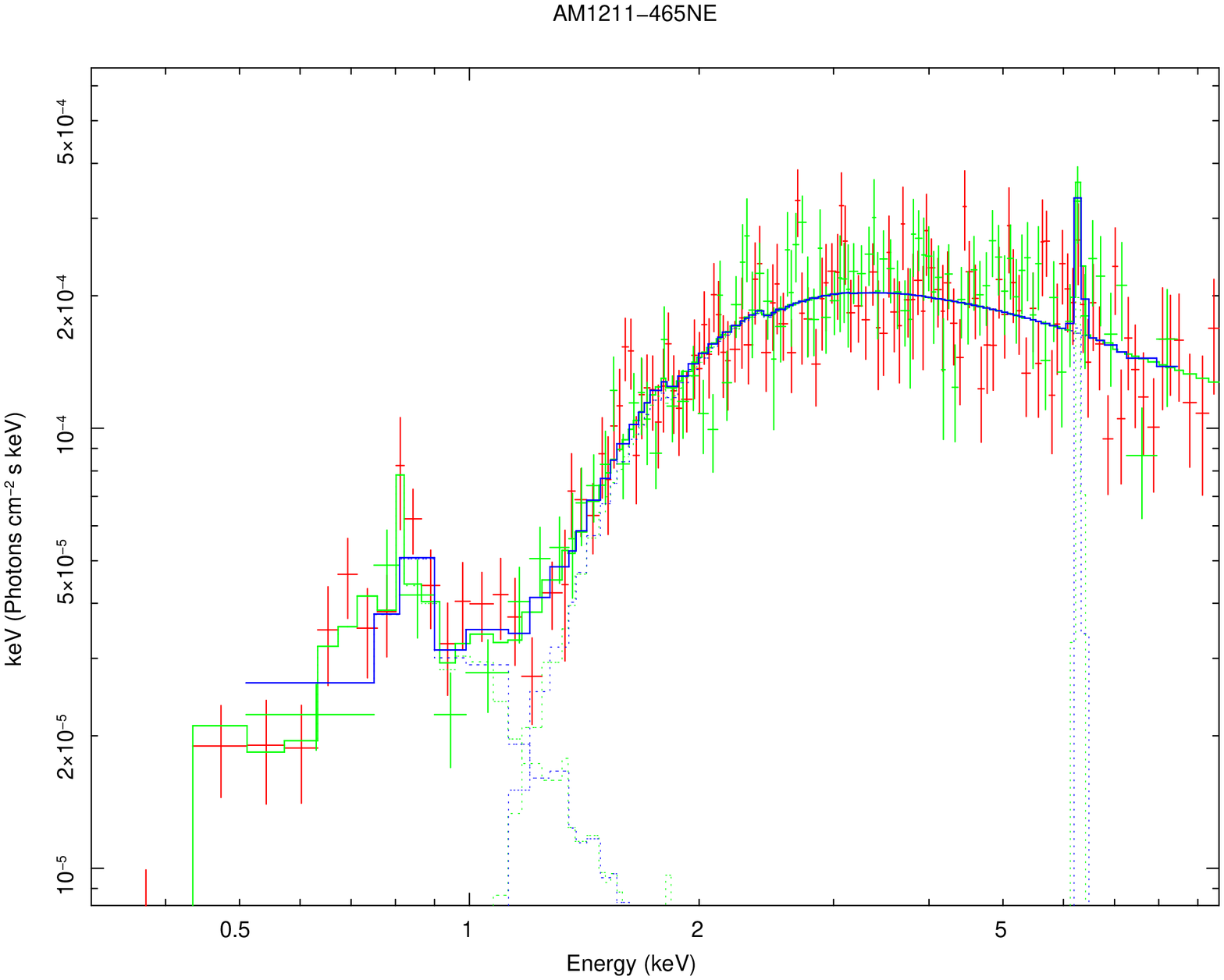}\\

\caption{ Observed  spectrum,  best-fit  model,  and
residuals of {\it AM 1211-465-NE} on the left, and the unfolded spectrum
and  the model  in Ef(E) on the right. }

\label{fig:am1211ne_spectra}
\end{figure*}

\begin{figure*}

\includegraphics*[width=60mm,angle=-90]{am1211SW_spec.ps}\includegraphics*[width=60mm,angle=-90]{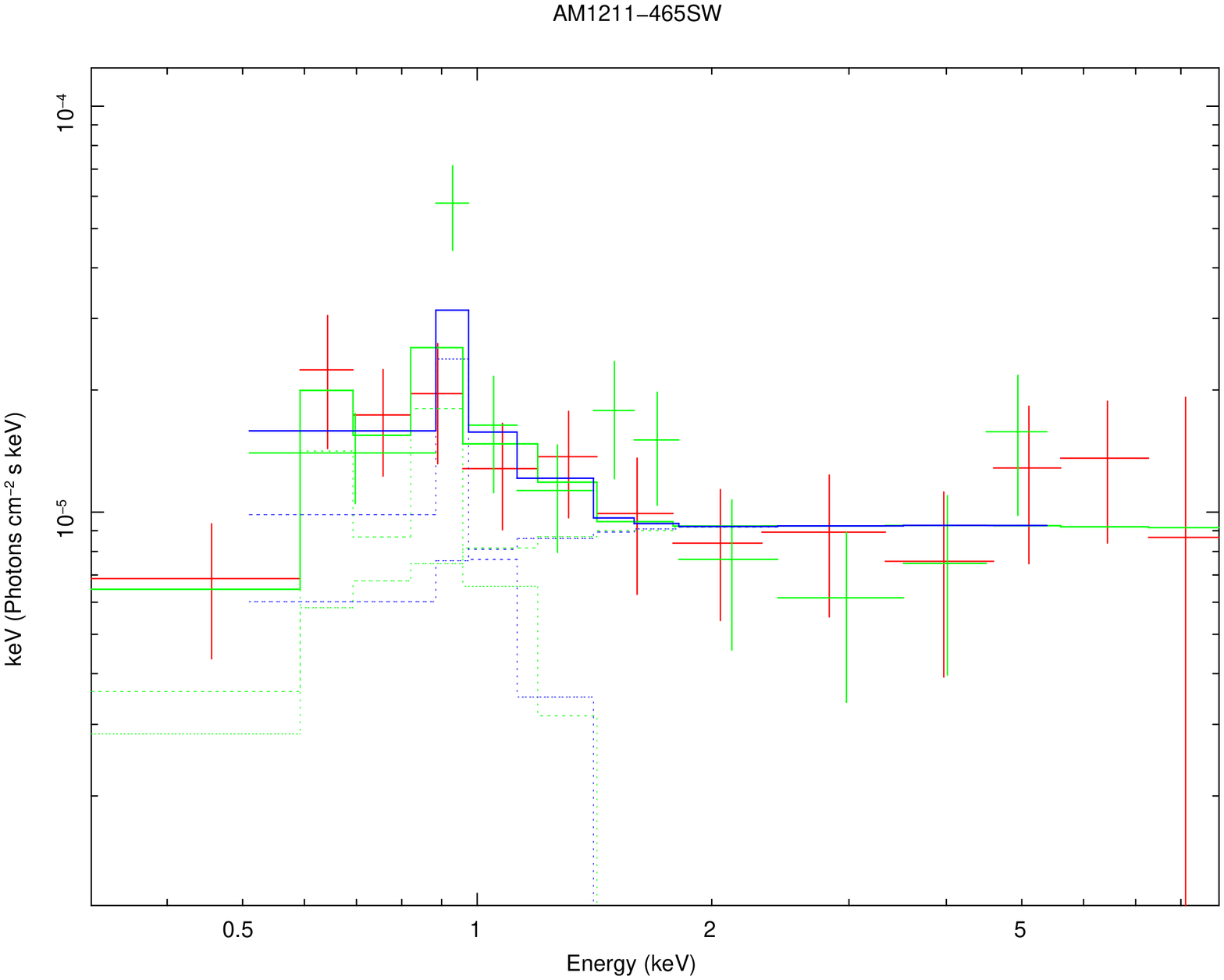}\\

\caption{Observed  spectrum,  best-fit model, and  residuals of  {\it  AM 1211-465-SW}  on  the  left and  the
unfolded spectrum and the model in Ef(E) on the right. }

\label{fig:am1211sw_spectra}
\end{figure*}

\begin{figure*}
\includegraphics*[width=60mm,angle=-90]{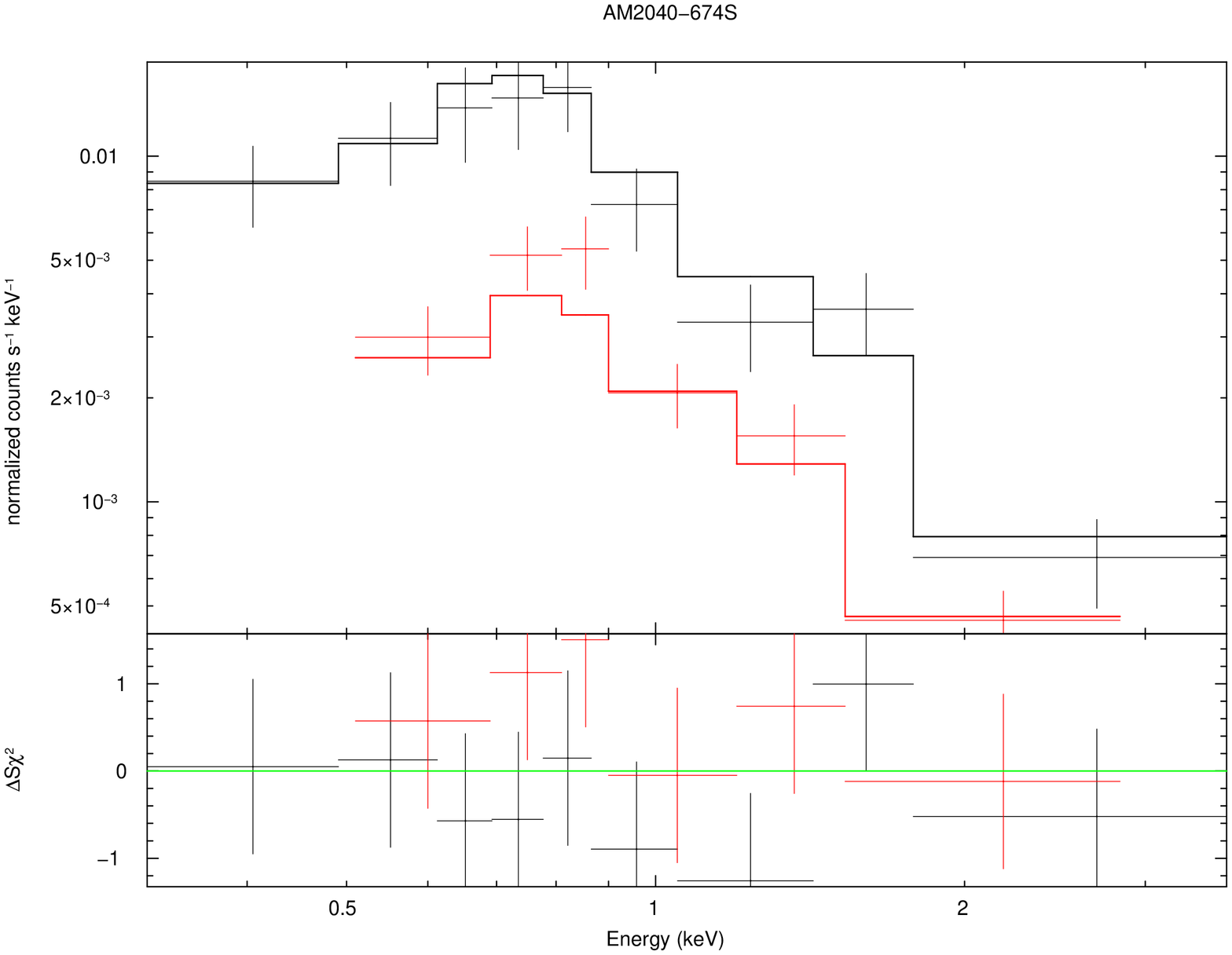}\includegraphics*[width=60mm,angle=-90]{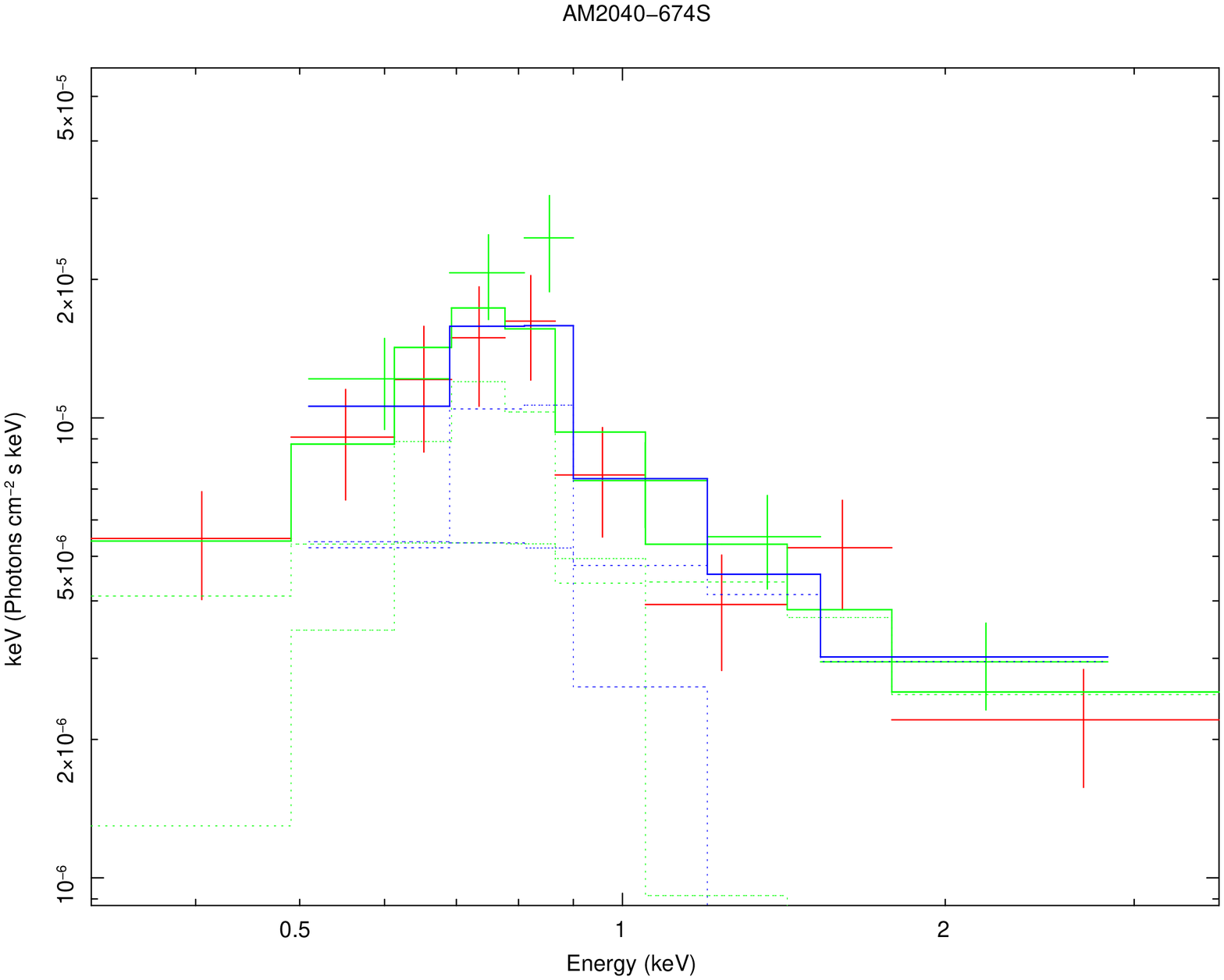}\\

\caption{Observed  spectrum,  best-fit  model,  and
residuals of {\it AM 2040-674-S} on the left and the unfolded spectrum
and  the model  in Ef(E) on the right. }

\label{fig:am2040_spectra}
\end{figure*}

\section{Discussion}

It is  accepted  that merging  activity induces starforming
episodes.  However,  we still do  not fully understand  the connection
between  the origin of  the energy  budget and  the properties  of the
involved galaxies, such as the pair  separation, the shapes of the galaxies,
or a  possible presence  of an  AGN. In this  section,  we  present the
X-ray  properties of  the three  interacting systems  and  discuss the
nature  of  their  emission.  We  also consider  the  results  in  the
framework  of  the scenario  of  black hole  activation in  merging
systems (Mortlock et al. 1999).  Finally, we compare our results with
the very few other cases where nearby binary AGN were detected.

\subsection{Starburst versus AGN as the origin of the nuclear X-ray emission}

Among the six studied  sources, only AM1211-465NE can be unambiguously
classified  as an  AGN.  The  imaging analysis  supports  a point-like
nature.     The     X-ray    luminosity    in     the    hard    band,
1.94$^{+0.11}_{-0.15}\times10^{42}$  ergs/s, is  difficult  to explain
with only star-formation  activity. The value found for  the power law
index, $1.63\pm0.10$,  also agrees well  with the mean  values observed
for AGN.  Interestingly,  the soft excess is preferably  fitted with a
{\it   mekal}    model,   but    with   a   very    low   metallicity,
0.030$^{+0.03}_{-0.018}$,  or a  {\it  bremsstrahlung} component.   In
general, starforming regions tend  to be  fitted with higher
metallicities (e.g. Strickland et al.  2004), due to the enrichment of
the environment by Type-II SN,  (e.g.  Arnett 1995). However, low
metallicity values have been measured (Strickland \& Stevens, 2000) in
starforming regions. This low value is  probably due to an  artifact of fitting
the multi-phase plasma emission  with single-temperature models.  The
measurement of the equivalent hydrogen column of the neutral absorbing
material,  $2.2\pm0.2\times10^{22}$   cm$^{-2}$,  is  compatible  with
typical values found for Seyfert  2 galaxies. If this neutral material
is  located in  the AGN  itself, its  origin could  be related  to the
shielding torus.   Another possibility is that the  merging process is
causing  gas  inflows  that are effectively  obscuring  the  nuclear
emission.  This highly  obscuring material could be the  reason for the
classification of AM1211-465NE as a  HII galaxy (Corbett et al.  2003;
Sekiguchi \&  Wolstencroft 1992) using optical and  IR data.  Finally,
the  detection  (98.8\% according  to  the  F-test)  of a  neutral  Fe
K$\alpha$  emission  line reinforces  the  AGN  origin  of the  nuclear
emission of AM1211-465NE. Allowing the energy of line to vary did
not improve  the goodness of the  fit. However we measured  a value of
$6.34^{+0.14}_{-0.2}$ keV, incompatible with the ionized iron emission
line typically at 6.7~keV. In fact, if the energy of the line is fixed
to this value,  the addition of this component  does not improve the fit.

On  the other  hand,  the spatial  and  spectral results  on the  both
components of the AM2040-674 pair and AM0707-273W indicates that their
X-ray   emission   can  be   easily   explained  with   star-formation
processes. The image of AM2040-  674N shows a diffuse emission visible
only in  the soft band.   Too few counts  are detected for  a spectral
analysis to be  possible, but upper limits on the  soft and hard X-ray
fluxes    were   derived    by   assuming    an    absorbed   (n$_{\rm
{H}}=$1$\times10^{21}$~cm$^{-2}$)  power-law ($\Gamma=1.8$)  emission
model   (see  Table~\ref{tab:lum}).   For   the  other   two  objects,
AM0707-273W and AM2040-67S, the hard and soft X-ray luminosities are
lower than 10$^{40}$  erg/s, in the range of  HII and normal galaxies.
Moreover,  both X-ray  spectra can  be  fitted with  a double  thermal
component with reasonable values  for the temperatures, although these
fits  are significantly  worse than  power law  plus  thermal emission
models.   We  derived  the  star-formation  rate (SFR)  of  these  two
galaxies using the empirical relationship with the 2-10~keV luminosity
of  Grimm et  al.   (2003): $L_{2-10\,keV}=6.7\times10^{39}$SFR.   The
values   obtained    are   1.2$M_\odot$/yr   for    AM0707-273W   and
6.0$M_\odot$/yr for AM2040-674S. An upper limit, SFR$<6.4M_\odot$/yr,
was calculated for AM2040-674N.  According to Kennicutt (1998),
the   FIR  luminosities  derived   from  these   values  of   the  SFR
(SFR$(>0.1{\rm    M}_{\odot})=\frac{{\rm   L_{FIR}}}{5.8\times10^9{\rm
L}_\odot}\,{\rm M_{\odot}yr^{-1}}$) are within the same order of magnitude
or even lower than the  observed with IRAS. In particular, the L$_{\rm
FIR}$      measured     for      the     AM~2040-674      pair     is
$2.6\times10^{-10}$~ergcm$^{-1}$s$^{-1}$ and the L$_{\rm FIR}$ derived
from         the         SFR         of        AM2040-674S         is
$0.6\times10^{-10}$~ergcm$^{-1}$s$^{-1}$.   For  the AM0707-273,  the
observed  L$_{\rm   FIR}$  is  $8\times10^{-10}$~ergcm$^{-1}$s$^{-1}$,
while  the   L$_{\rm  FIR}$  derived   from  the  calculated   SFR  is
$3.5\times10^{-10}$~ergcm$^{-1}$s$^{-1}$    (see   below    for   the
contribution of AM0707-273E).

The spectral properties  of the E member of  AM0707-273 are compatible
with both an AGN and  powerful starburst origin.  The imaging analysis
favors the  second possibility because the radial  profile suggests an
extended   emission.    The   temperature   of   the   mekal   plasma,
kT=$0.143^{+0.07}_{-0.002}$~keV, is compatible with the typical values
obtained for  AGN.  On  the other  hand, the index  of the  power law,
1.2$\pm0.2$  is  lower than  the  typical  values  of AGN.   Likewise,
although     the    luminosity     measured     for    the     galaxy,
L$_{\rm{0.3-10\,keV}}\sim2\times10^{42}$  erg/s is high  compared with
the X-ray  luminosities of normal or  HII galaxies, it is  in the soft
band where its  contribution is higher. This high  luminosity could be
an artifact  of the  high absorbing column  associated to  the thermal
component,  n$_{\rm  {H}}=1$$\times10^{22}$~cm$^{-2}$.   In  fact,  we
calculated         a         lower         absorption,         n$_{\rm
{H}}\sim2$$\times10^{21}$~cm$^{-2}$,   using   the  H$\alpha$/H$\beta$
ratio obtained  from Sekiguchi  \& Wolstencroft (1992).   We therefore
fixed the index  of the power law to 1.8  (compatible with mean values
of $\Gamma$ found in both AGN and starforming galaxy spectra), leaving
the rest of the parameters  free.  Although the value of n$_{\rm {H}}$
associated  to   the  thermal  component  decreased,   the  soft  band
luminosity,  L$_{\rm{0.5-2\,keV}}\sim7\times10^{41}$  erg/s, is  still
high.  However, in  the hard band, the measured  luminosity is only on
the order  of 10$^{40}$ erg/s.   Although it is  a luminous
object in the X-ray band, its  nature could also be compatible with
intense starburst activity.  If this is the case, the SFR derived from
the 2-10~keV  luminosity of  the source (Grimm  et al. 2003)  would be
2.2$M_\odot$/yr.

Finally,  the  results  of  the  analysis  of  AM1211-465SW  are  more
ambiguous. The  imaging analysis indicates a  point-like emission. The
power law is very flat, $\Gamma = 1.0\pm0.5$, and the total luminosity
is  L$_{\rm{0.3-10\,keV}}\sim4\times10^{42}$ erg/s,  even  higher than
the value measured  for AM1211-465NE.  We also tried  to fix the index
of the power law to 1.8.  The value of n$_{\rm {H}}$ associated to the
thermal component decreased slightly, but the luminosity in the 0.3-10
keV band remains unvaried.  In  the hard band, the measured luminosity
is   also  high,   $\sim  1\times10^{41}$   erg/s,  higher   than  the
luminosities observed for HII and LINER but lower than typical Seyfert
2 luminosities. One possible explanation would be that the source is a
highly absorbed  AGN with an  equivalent hydrogen column  of 10$^{24}$
cm$^{-2}$ or more, which would  explain the low value measured for the
index of  the power law.  We  studied the starforming  activity of the
source.    Based  on  the   hard  luminosity,   the  SFR   derived  is
13$M_\odot$/yr.  The SFR of  AM1211-465SW is  higher than  the values
derived with the  same method for the classified  {\it pure} starburst
galaxies, i.e. AM0707-273W and  both members of AM2040-674, pointing
to the possibly LLAGN nature of  this source. In any case, there is no
strong evidence of the presence of a luminous AGN in AM1211-465SW and
the X-ray  activity is very likely  due to a low-luminosity  AGN or a
very intense starburst activity.

\subsection{Activation of quiescent black holes}

It is  widely accepted that  optical spectroscopy alone can  provide a
misleading  classification of galactic  nuclei.  Indeed,  an extensive
study performed  with \chandra\ (Maiolino  et al. 2003)  revealed that
the  density  of   AGN  is  twice  the  one   estimated  from  optical
spectroscopic surveys.  This  misclassification in the optical studies
is   associated   with   their   high  sensitivity   to   obscuration.
Interestingly, in merging systems  where the presence of obscuring gas
is expected, the  detection of elusive AGN is about 50\% of
the studied systems  in the X-ray band.  The  most outstanding example
of a  binary AGN  was discovered  by Komossa et  al.  (2003)  based on
\chandra\ observation  of the ultraluminous galaxy  NGC6240.  Ballo et
al.  (2004) suggest that both galaxies of the merger Arp 299 host an
AGN-type nucleus.  Both  galaxies were spectroscopically classified as
starforming galaxies or LINER based  on optical and mid-IR data. More
recently,  Guainazzi   et  al.   (2005)   observed  ESO509-IG066  with
\xmm. The merger system hosts  two AGN with X-ray luminosities of $\sim
10^{43}$ erg/s.  One of the  galaxies of the pair was misclassified as
an  HII galaxy  based on  optical  and IR  observations (Sekiguchi  \&
Wolstencroft  1992).  In  this work,  the X-ray  properties of  the SW
component  of the  AM1211-465 pair  suggest  an AGN  nature for  its
nucleus.  If this is the  case, AM1211-465 would be the fourth example
of an AGN pair in a merging system.

Exhaustive  studies of known  quasar pairs  reveal that  the galaxies,
when merging, could become active  (Mortlock et al.  1999).  The model
suggests  that  when the  galaxies  reach  a  certain distance,  tidal
interactions  cause  gas  flowing  into  the  cores  of  the  galaxies
switching on the black holes.   Taking into account 16 good candidates
to binary  quasars, Mortlock et al.  (1999)  estimated this activation
distance to  be in the  range between 50  to 100 kpc. When  the merger
becomes more stable, at typical distances of 10 kpc, the inflow of gas
would  cease and  the  back  holes become  quiescent  again.  The
activation distance range estimated in Mortlock et al. (1999) is based
on  projected physical  separation that considers  different cosmological
models.  The  authors therefore expect that the  binary quasars exist
in very distorted hosts and not in relaxed mergers.  The assessment of
the  critical  distance is  important  for    understanding  the
processes  that  lead to  nuclear  activity  in interacting  galaxies.
According  to this  model, the  activation the  the two  nuclear black
holes can  only occur  in encounters of  galaxies with  similar sizes,
since the interaction between dwarf and large galaxies would result in
a single AGN.  However, only a small fraction of quasars have a second
nearby   quasar  because  in   general  the   pairs  are   in  reality
gravitational lenses  or chance  alignments.  Therefore, only  a small
number  of  true  (or  good)  candidates of  quasar  pairs  have  been
identified so far and most of them have high redshifts (z=1-2.5).

As a  matter of fact, X-ray  studies are efficient  at discovering hidden
pairs of AGN.  Due to a  selection effect, all AGN pairs discovered in
the X-ray band  are nearby objects.  In these  cases, we are therefore
able to determine  the properties of the host  galaxies, which in many
of  the distant  pairs are  not even  detected.  The  three  AGN pairs
detected  in  the  X-rays,  NGC6240,  Arp299,  and  ESO509-IG066,  have
an apparent  separation distance  in the  range of  1.4-13 kpc,  lower or
close  to the  minimum  value  of the  activation  distance, i.e.   10
kpc. Only the host galaxies  of the two former mergers present clearly
disrupted  shapes.  

In  this  work,  we  have  detected  a  possibly  double  AGN  system,
AM1211-465. The apparent  separation of the galaxies is  98 kpc, so in
this  case, close  to  the  upper limit  of  the activation  distance.
Finally, contrary to the Mortlock hypothesis, no sign of disruption is
observed  in the  X-ray images  of  any of  the two  host galaxies  of
AM1211-465.

\subsection {Comparison with other merging galaxies}

Excluding AM1211-465NE and AM1211-465SW, the derived properties of the
rest  of  the  objects  are  compatible  with  those  found  in  other
interacting systems where  no evidence of AGN has  been found (Jenkins
et al. 2005 and references therein). In general, the X-ray spectrum of
the interacting galaxies is characterized  by a power law and a single
or a  double thermal  component present in  the soft energy  band. The
X-ray    (0.3-10    keV)     luminosities    are    in    the    range
(0.2-5)$\times10^{41}$ erg/s and  are explained by intense starforming
activity  triggered  by   the  merging  process.  The  high-resolution
\chandra\ X-ray images show that many of the objects host luminous and
ultraluminous X-ray  sources (ULX).   Brassington et al.   (2005) find
seven ULX  out of 16 point-like  sources spread in  the merging system
Arp270, and similar results are also observed in other mergers such as
NGC7714/NGC7715 (Smith,  Struck \& Nowak,  2005) and Arp299  (Ballo et
al. 2004; Zezas, Ward \&  Murray 2003). The limited spatial resolution
of  our \xmm\  observation did  not  allow the  detection of  possibly
off-nuclear point-like X-ray sources.  Higher resolution images would
help  to unambiguously  determine  where the  hard  X-ray emission  of
AM1211-465SW and AM0707-273E originate.   Jenkins et al.  (2005) find
that the \xmm\ spectrum of NGC7771  is compatible with being due to an
AGN,  although the  evidence  is not  strong.   Its companion  galaxy,
NGC7770, is classified  as a Starburst galaxy. In  this sense, even if
only  the  NE  component of  the  AM1211-465  system  is an  AGN,  the
detection  favors the  role  of fueling  quiescent supermassive  black
holes  in  merging systems  and  the  importance  of systematic  X-ray
studies to unveil the AGN, obscured by the merging process.

\section{Conclusions}

In  this  paper we  have  presented  the  X-ray imaging  and  spectral
analysis   of  three  pairs   of  interacting   galaxies,  AM0707-273,
AM1211-465,  and  AM2040-674, observed  with  \xmm.  Six galaxies  were
detected and the  \xmm\ spatial resolution allowed us  to isolate each
 member of the pairs.

The images  of the  galaxies show nuclear  and extended  diffuse X-ray
emission for  all pairs. All  the galaxies, except AM2040-675S, were
detected  in the  hard band.   No evidence  of  off-nuclear point-like
sources were observed in any  of the galaxy pairs. However, the coarse
spatial resolution  of the \epic\  cameras on-board \xmm\  prevents us
from  making  strong  statements  on  this point.   Hints  of  merging
processes were observed  for AM0707-273 and AM1211-465 in  the form of
intergalactic soft X-ray emitting  gas. The apparent separation of the
two AM2040-674  galaxies does not  allow us to disentangle  the emission
of the host galaxies from the emission due to possible disruption of
the merging process.

For five  (the exception being  AM2140-674S) out of the  six galaxies,
enough  counts  were  detected   to  allow  spectrum  extractions  and
analysis. All the spectra analyzed  are complex, and two components are
needed to explain the 0.3-10 keV emission.  The best-fit model for all
the sources  consists of a power  law, accounting for the  bulk of the
hard emission  and a thermal  mekal or bremsstrahlung  component that
explains    the   ubiquitously   observed    excess   in    the   soft
band. 
The luminosities  in the 0.3-10 keV band range from (0.4-40)$\times10^{41}$ erg/s.

Optical and IR spectroscopic studies previously classified the nuclear
activity   of  all   galaxies  as   powered  by   intense  starforming
processes. The \xmm\  data allowed us to unambiguously  unveil the AGN
nature of AM1211-465NE,  misclassified in the optical due  to its high
obscuration.   The  spectrum   shows  the  marginal  presence  (98.8\%
according to the F-test)  of the neutral FeK$\alpha$ line, reinforcing
the AGN nature of the source. The spectral properties of its companion
galaxy, AM1211-465SW, are compatible with  the presence of an AGN.  If
this is the  case, AM1211-465 would be the fourth  example of a binary
AGN discovered in X-rays. AM0707-273E could also host a low-luminosity
AGN, although the evidence is  not strong. The X-ray classification of
the three remaining  sources, AM0707-273W and both  members of AM2040-674,
as HII galaxies  agrees with the results  of other studies in
optical/IR.

Our  results  do  not  contradict  the theory  of  the  activation  of
quiescent  black   holes  through  the  gas   accretion  triggered  by
encounters of galaxies (Mortlock et  al. 1999). Although, we have only
found signatures  of double AGN  in one galaxy pair  (AM1211-465), its
projected members'  distance ($\simeq  100$~kpc) is comparable  to the
AGN activation radius. Our results  also point to the distance between
galaxies in  merging systems not being the  only determining parameter
in the black hole activation. However, this type of analysis shows the
importance of X-ray studies of galaxy pairs for accurately determining
the  nature of  their nuclei,  in particular  in those  suffering high
absorption.  Systematic  and high-resolution X-ray surveys  of galaxy
pairs are important for furthering our understanding of the mechanisms
that originate the nuclear activity of  AGN, as well as for a complete
demography of them.

\acknowledgements{  The  authors would  like  to  thank the  anonymous
referee for the very  useful comments that significantly improved the
work. The authors would also want to thank Miguel Mass-Hess and H\'ector
Ot\'{\i}-Floranes for helpful discussions. EJB and MSL acknowledge funding from
the Spanish MEC grant AYA2004-08260-C03-03. DRG acknowledges support by the
Mexican research council (CONACyT) under the grant 49942.}\\


\begin{thebibliography}{9}

%
%

\bibitem[]{} 
Arnaud, K.~A.,\ 1996, {ASP Conf.Ser. Vol. 101, Astronomical  Data Analyis and Systems, 17}


\bibitem[]{} 
Arnett, D.\ 1995, \araa, 33,  115 


\bibitem[]{} 
Avni, Y.\ 1976, \apj, 210, 642 


\bibitem[]{} 
Ballo, L., Braito, V.,  Della Ceca, R., Maraschi, L., Tavecchio, F., \& Dadina, M.\ 2004, \apj, 
600, 634 


\bibitem[]{} 
Bennett, C.~L., et al.\  2003, \apjs, 148, 1 


\bibitem[]{} 
Brassington, N.~J.,  Read, A.~M., \& Ponman, T.~J.\ 2005, \mnras, 360, 801 

\bibitem[]{} 
Cappi, M., et al.\ 1999,  \aap, 350, 777 

\bibitem[]{} 
Corbett, E.~A., et al.\  2003, \apj, 583, 670 



\bibitem[]{} 
Dickey, J.~M., \&  Lockman, F.~J.\ 1990, \araa, 28, 215 


\bibitem[]{}
Fouque, P., Durand, N., Bottinelli, L., Gouguenheim, L. \& Paturel, G., 1992, Catalogue of Optical Radial Velocities, vol. 1 p. 1 

\bibitem[]{} 
Gabriel, C., et al. 2004, ASP Conf.~Ser.~314: Astronomical Data Analysis Software and Systems 
(ADASS) XIII, 314, 759 

\bibitem[]{} 
Gonzalez-Martin, O., Masegosa, J., Marquez, I., Guerrero, M. A.\& Dutzin-Hacyan, 2006, D. astro-ph/060562


\bibitem[]{} 
Grimm, H.-J., Gilfanov,  M., \& Sunyaev, R.\ 2003, \mnras, 339, 793 



\bibitem[]{} 
Guainazzi, M.,  Piconcelli, E., Jim{\'e}nez-Bail{\'o}n, E., \& Matt, G.\ 2005, \aap, 429, 
L9 


\bibitem[]{} 
Hennawi, J.~F., et al.\  2006, \aj, 131, 1 


\bibitem[]{} 
Hewett, P.~C., Foltz,  C.~B., Harding, M.~E., \& Lewis, G.~F.\ 1998, \aj, 115, 383 


\bibitem[]{} 
Ho, L.~C., et al.\ 2001, \apjl, 549, L51 


\bibitem[]{}
Jansen F., Lumb D., Altieri B. et al. 2001, A\&A, 365, L1


\bibitem[]{} 
Jenkins, L.~P., Roberts, T.~P., Ward, M.~J., \& Zezas, A.\ 2005, \mnras, 357, 109 


\bibitem[]{} 
Jenkins, L.~P.,  Roberts, T.~P., Ward, M.~J., \& Zezas, A.\ 2004, \mnras, 352, 1335 



\bibitem[Jim{\'e}nez-Bail{\'o}n   et   al.(2005)]{2005A&A...435..449J}
Jim{\'e}nez-Bail{\'o}n, E.,  Piconcelli, E., Guainazzi,  M., Schartel,
N.,  Rodr{\'{\i}}guez-Pascual, P.~M.,  \&  Santos-Lle{\'o}, M.\  2005,
\aap, 435, 449


\bibitem[]{} 
Jogee S., Scoville N., Kenney J.~D.~P., 
2005, ApJ, 630, 837 



\bibitem[]{}
Kendall, M.G., \& Stuart, A., 1973, The advanced Theory of Statistics Vol. 2, Hafner, New York, Section 19.26, page 97

\bibitem[Kennicutt(1998)]{1998ApJ...498..541K} Kennicutt, R.~C., Jr.\ 1998, 
\apj, 498, 541 

\bibitem[]{} 
Kochanek, C.~S.,  Falco, E.~E., \& Mu{\~n}oz, J.~A.\ 1999, \apj, 510, 590 




\bibitem[)]{} 
Komossa, S., Burwitz, V., Hasinger, G., Predehl, P., Kaastra, J.~S., \& Ikebe, Y.\ 2003, \apjl, 
582, L15 


\bibitem[]{} 
Kormendy, J., et al.\ 1997, \apjl, 482, L139 


\bibitem[]{} 
Maiolino, R., et al.\  2003, \mnras, 344, L59 

\bibitem[]{}
 Morrison, R., \&  McCammon, D.\ 1983, \apj, 270, 119 


\bibitem[]{} 
Mortlock, D.~J.,  Webster, R.~L., \& Francis, P.~J.\ 1999, \mnras, 309, 836 

\bibitem[]{}
Persic, M., \& Rephaeli, Y.\ 2002, \aap, 382, 843 

\bibitem[]{} 
Piconcelli, E., 
Jimenez-Bail{\'o}n, E., Guainazzi, M., Schartel, N., 
Rodr{\'{\i}}guez-Pascual, P.~M., \& Santos-Lle{\'o}, M.\ 2004, \mnras, 351, 
161 


\bibitem[]{} 
Sekiguchi,  K., \& Wolstencroft, R.~D.\ 1992, \mnras, 255, 581 


\bibitem[]{} 
Shlosman I., 2005, AIPC, 783, 223 



\bibitem[]{} 
Smith, B.~J., Struck, C., \& Nowak, M.~A.\ 2005, \aj, 129, 1350 

\bibitem[]{} 
Strauss, M.~A., Huchra, J.~P., Davis, M., Yahil, A., Fisher, K.~B., \& Tonry, J.\ 1992, \apjs, 83, 
29 


\bibitem[]{} 
Strickland, D.~K.,  Heckman, T.~M., Colbert, E.~J.~M., Hoopes, C.~G., \& Weaver, K.~A.\ 2004, 
\apjs, 151, 193 


\bibitem[]{} 
Theureau, G., et al.\ 2005, \aap, 430, 373 

\bibitem[]{} 
Wong, O.~I., et al.\ 2006, \mnras, 371, 1855 




\bibitem[]{} 
Zezas, A., Ward, M.~J., \& Murray, S.~S.\ 2003, \apjl, 594, L31 



\end{thebibliography}
\end{document}